\documentclass[prb,aps,amssym,nofootinbib,floatfix,eqsecnum]{revtex4} 
\usepackage{graphicx,natbib}
\newcommand{\ep}{\tilde \epsilon_0}
\newcommand{\be}{\begin{equation}}
\newcommand{\ee}{\end{equation}}
\newcommand{\bea}{\begin{eqnarray}}
\newcommand{\eea}{\end{eqnarray}}
\begin{document}
\title{Non-Hermitian Luttinger liquids and flux line pinning in 
planar superconductors}
\author{Ian Affleck,$^{1,2}$ Walter Hofstetter,$^{3}$ 
David R. Nelson$^4$ and Ulrich Schollw\"ock$^5$}
\affiliation{
$^1$Department of Physics, Boston University, Boston, MA 02215, USA \\
$^2$Department of Physics \& Astronomy, University of British Columbia, 
Vancouver, B.C., Canada, V6T 1Z1\\\
$^3$Physics Department, Massachusetts Institute of Technology, 
Cambridge, MA 02139, USA\\
$^4$Lyman Laboratory, Harvard University, Cambridge, MA 02138, USA \\
$^5$Institute of Theoretical Physics C, RWTH Aachen, D-52056 Aachen,
 Germany}
\date{\today}
\begin{abstract}
    As a model of thermally excited flux liquids connected by a weak
    link, we study the effect of a single line defect on vortex
    filaments oriented parallel to the surface of a thin planar
    superconductor.  This problem can be mapped onto the physics of a
    Luttinger liquid of interacting bosons in 1 spatial 
    dimension with a point impurity.  When the applied magnetic field
    is {\it{tilted}} relative to the line defect, the corresponding
    quantum boson Hamiltonian is \emph{non-Hermitian}.   We analyze this
    problem using a combination of analytic and numerical (density
    matrix renormalization group) methods, uncovering a delicate
    interplay between enhancement of pinning due to Luttinger liquid
    effects and depinning due to the tilted magnetic field.
    Interactions allow a single columnar defect to be very effective
    in suppressing vortex tilt when the Luttinger liquid parameter
    $g\leq 1.$        
\end{abstract}
\pacs{}
\maketitle

\section{Introduction}
The past fifteen years have seen much work on the
statistical mechanics and dynamics of thermally excited vortices
in Type II high temperature 
superconductors.\cite{bishop92,blatter94,tauber97,nattermann00}
The
competition between interactions, pinning and thermal fluctuations
gives rise to a wide range of novel phenomena, including a first
order melting transition of the Abrikosov flux lattice into an
entangled liquid of vortex filaments,\cite{nelson88} a proposal for a highly
disordered {\it{vortex}} glass dominated by point pinning,\cite{fisher89a}  a
theory of a {\it{Bose}} glass phase with vortices strongly pinned
to columnar defects,\cite{nelson92} and a distinct {\it{Bragg}} glass where
point disorder converts Bragg peaks associated with crystalline
order into power law singularities.\cite{giamarchi94}

Much progress can be made on these problems using a classical continuum 
elastic theory.\cite{blatter94} An alternative approach, useful 
for treating parallel columnar defects, is to regard 
each vortex as an imaginary-time world line of a boson  
in a Feynman path integral, corresponding 
to a quantum theory of interacting bosons.\cite{nelson88,fisher89b}
  Here the imaginary 
time direction $\tau$ is parallel to the columnar defects. 
A hydrodynamic treatment of this quantum model leads naturally 
to the same continuum elastic theory, where the classical 
free energy is now a classical action. 
If  the direction of the external magnetic field does
{\it{not}} coincide with that of columnar defects, it is
convenient to separate the transverse component of the field
$H_\bot$ from the parallel component $H_{||}$ along ${\bf{\hat
\tau}}.$ When $H_\bot << H_{||}$, the transverse component $H_\bot$
plays the role of a constant imaginary vector potential for the
bosons.\cite{nelson92,hatano97}
   The corresponding fictitious quantum Hamiltonian
is {\it{non-Hermitian}}, with new and interesting properties.
Stimulated by vortex physics, there has been considerable work
on non-Hermitian models of non-interacting bosons in a constant
imaginary vector potential,  $h\propto H_\bot$, and a disordered
site-diagonal pinning.\cite{efetov97,brouwer97,mudry98}
 Since the Hamiltonian is non-Hermitian,
the energy eigenvalues can be either real or complex.  As
discussed in [\onlinecite{hatano97}], all states with complex eigenvalues are
extended, whereas those with real eigenvalues are usually
localized.

Less is known about non-Hermitian models {\it{with}}
interactions.\cite{nelson92,hatano97} A disordered array of parallel columnar
defects leads to a strongly pinned low temperature Bose glass
phase. For $h$ less than a critical value $h_c$,  this phase
exhibits a ``transverse Meissner effect", such that the vortex
filaments remain pinned to the columns even though the external
field is tilted away from the column direction. Although the
transverse Meissner effect has now been observed in many high
$T_c$ superconductors with correlated disorder, recent
measurements\cite{phuoc02}
 of the flux flow resistivity for $h_\bot \gtrsim h_c$
  disagree with a simple theory which assume that vortices tilt
from column to column via kink excitations.\cite{hwa93} Interactions
combined with vortex pinning for a {\it{periodic}} array of
columnar defects were studied using a non-Hermitian boson Hubbard
model in [\onlinecite{lehrer98}]. Here, tilting the field drives a transition
out of a low temperature ``Mott insulator" phase (periodic array
of vortices attached to the columns with an energy gap)  into a
``superfluid" phase, i.e., an entangled flux liquid.  The
corresponding non-Hermitian boson Hubbard model with site-diagonal
disorder was studied in $(1 + 1)$-dimensions using a
Hartree-Bogoliubov approximation in [\onlinecite{kim01}].

\begin{figure}
\begin{center}
\includegraphics[width=0.45\linewidth]{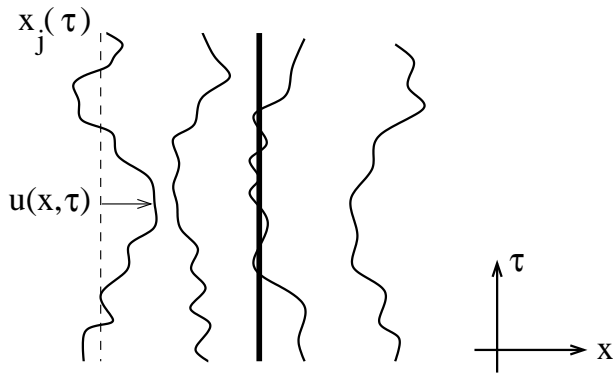}
\caption{ Vortices with a single columnar pin (heavy vertical line) and parallel
magnetic field with  
displacement field $u(x,\tau)$ defined.}
\label{fig1}
\end{center}
\end{figure}

In this paper, we study the effect of a {\it{single}}
columnar pin on the statistical mechanics of thermally fluctuating
vortex lines confined  in a thin, superconducting slab. (See Fig.
(\ref{fig1}).)  The external field can tilt away from the direction of the
column, leading to statistical physics controlled by 
a non-Hermitian quantum Hamiltonian.  As
discussed below, the physics is equivalent to a Luttinger liquid
of interacting bosons with a point impurity. Tilting the field
 introduces a constant imaginary vector potential.

This problem is interesting for a number of reasons.
 If the
average spacing between columnar defects is $d$, this is the
regime $H >> B_\phi = \phi_0/d^2$ where $B_\phi$ is the
``matching field'' and $\phi_0 \approx 2.07 \times 10^{-7}
{\rm G\, cm^2}$ is the flux quantum.  As emphasized by 
Radzihovsky,\cite{radzihovsky95}
 the large number of ``interstitial'' vortices between
columnar defects could be locally crystalline or melted into a
flux liquid. A third possibility is vortices in an immediate
``supersolid'' phase, which is both crystalline but nevertheless
entangled, due to a finite
concentration of line-like vacancy or interstitial defects.\cite{frey94}
By studying the response to a {\it single} columnar pin we 
can better understand the response in these phases to a dilute concentration of 
columnar defects. A dilute concentration of {\it twin planes}, 
a common occurance in bulk samples of YBCO, provides a related 
example of correlated pinning.

We examine here a similar regime in $(1 + 1)$- dimensions,
when only a single columnar pin is present. The feasibility of
studying vortex physics in samples which are effectively $(1 +
1)$-dimensional was demonstrated by Boll\'e et. al\cite{bolle99}, who used
micromechanical oscillators to track the entry of quantized vortex
filaments near $H_{c1}$ in a thin sample of the low $T_c$ 
superconductor ${\rm Nb Se_2}$.
The observed behavior could be described in the framework of
interacting vortex lines in a $(1 +1)$-dimensional random
potential representing the effects of point disorder.

Similar experiments might be possible on thin high - $T_c$
platelet samples, with temperatures high enough to allow vortex
interactions to screen out the effect of the point 
disorder.
The effect of a single columnar pin might be mimicked by gouging a
long straight scratch or notch on one side of a thin sample.\cite{zeldov}
A $(1 + 1)$ dimensional array of flux lines could be created by a field
$\vec H$, approximately parallel to the notch, and strong enough to
insert a single layer of flux lines into a sample with a thickness
given roughly by the London penetration depth.  As discussed
below, $(1 +1)$--dimensional arrays of vortex lines show algebraic decay
of both translational order and the boson order parameter. Thus,
experiments and theory on this $(1 + 1)$--dimensional problem might
give some insight into the effect of a dilute concentration of
columnar pins on a {{\it{supersolid}} phase in $(2 +
1)$-dimensions. Because of the long range correlations, a single
columnar defect can have a large effect on the flux liquid, similar to the
screening cloud surrounding a Kondo impurity in a metal.

A related problem in $(2 + 1)$-dimensions concerns the
effect of a single twin plane or grain boundary on vortex matter.
A {{\it{dense}} array of such planes parallel to the field
direction leads to a Bose glass phase at low temperatures and a
flux liquid at high temperatures.\cite{nelson92}  A single such plane should
have an interesting influence on the vortex matter which surrounds
it.  Consider the effect of a small tilt on the vortex
configurations.  The ``motion'' of the tilted vortex
configurations across the twin plane is an imaginary time version
of particle transport across a Josephson junction.  The
``transport'' process is likely to be quite different, depending
on whether the surrounding vortex matter is in a flux liquid,
vortex crystal or supersolid phase.  A single planar defect in $(2
+ 1)$-dimensions would also have a strong effect on the flux flow
resistivity in response to a current parallel to the plane leading to a
Lorentz force which is approximately perpendicular to it.  Motion
of the Bragg planes of the Abrikosov flux lattice across such a
defect is reminiscent of transport in materials with charge
density waves.\cite{lee}

The single defect, (1+1) dimensional physics problem is very tractible using 
the continuum elastic field theory approach. Its  
quantum hydrodynamic formulation corresponds 
to a single component ``Luttinger liquid''.\cite{Haldane}
The resulting field theory,
when the field is parallel to the defect, 
 is identical 
to one studied earlier\cite{Kane} 
in the context of interacting one-dimensional (spinless) fermions with 
a point defect and of a defect in an S=1/2
Heisenberg antiferromagnetic chain.\cite{Eggert}  We 
directly apply these results to the present situation. 
These results depend crucially on a dimensionless parameter, $g$
which controls all critical properties of the model.\cite{Kane} For instance 
the density correlations (with no defects) exhibit power law decay with an 
exponent $\eta =2g$. 
For interacting fermions $g<1$ corresponds to repulsive 
interactions and $g>1$ attractive interactions. In the 
bosonic case the dependence of $g$ on microscopic 
parameters is more subtle, as we discuss. When $g>1$ 
a defect is an irrelevant perturbation at long length scales,
in the renormalization group (RG) sense,  whereas 
when $g<1$ it is relevant, effectively flowing 
to infinity in the long wavelength limit.  As was 
 first suggested by DeGennes,\cite{deGennes} in the dilute 
limit our vortex problem is equivalent to free fermions, $g=1$
.\cite{deGennes,Pokrovskii,Coppersmith,Haldane} 
We show that a transverse magnetic field defines 
a characteristic length scale, $\propto  1/h$, which 
acts as an infrared cut off scale in the renormalization 
group flow equations. Thus even when $g<1$, a 
defect is ultimately an irrelevant perturbation 
at sufficiently long length scales.  

We also study a tight-binding version of the 
non-Hermitian 1D quantum model using the Density Matrix 
Renormalization Group (DMRG) method generalized to 
treat non-Hermitian Hamiltonians. This powerful method provides 
valuable checks on our RG arguments and more detailed 
information about the model.

We focus on the vortex density (i.e. magnetic field) oscillations 
set up by a single columnar pin as well as on the transverse 
Meissner effect. These correspond, respectively, 
 to generalized Friedel oscillations 
and an imaginary current in the quantum problem. While 
these Friedel oscillations have power-law decay for 
a parallel field, they decay exponentially when $h\neq 0$. 
The imaginary current can be expressed in terms of a 
``pinning number'', $N_p$, which measures how many 
bosons are stuck in the vicinity of the defect at 
any given ``time''. We find that $N_p$ can diverge as $h\to 0$, 
with  different critical behavior in the  
cases $g\leq 1$ and $g>1$. 

A real sample would always contains some density of point 
defects in addition to one or more line defect. We show 
that any finite density of point defects alters 
the critical behavior associated with the line defect 
at sufficiently large length scales. Thus to observe 
the critical behavior caused by a single line defect 
it will be neccessary to have sufficiently clean samples. 

In the next section we briefly review the classical continuum elastic 
theory of fluctuating lines and the approach based 
on mapping into a quantum model.  We also 
discuss the value of $g$, showing that 
it goes to $1$ at low densities and $0$ at higher densities. We note
that $g$ need not be a monotonic function of density for vortex arrays.  It is 
possible that $g$ passes through unity at some finite density as well. 
In Sec. III we study the $g=1$ case, which occurs 
in the dilute limit, 
 by exploiting 
the correspondence to non-interacting fermions.
 In Sec. IV we study general 
values of $g$ using renormalization group and numerical methods, extending 
earlier results\cite{Kane,Eggert} to the non-Hermitian case.
Sec. V contains a discussion of point defects. Sec. VI 
contains our conclusions.

Appendix A derives a result for interacting 
bosons in one dimension of general applicability. While 
it is well known that $g\to 1$ in the dilute limit, corresponding 
to free fermions, we study the leading correction to 
this behavior at small finite density, $n_0$.  We 
find that the result can be conveniently expressed 
in terms of the even channel scattering length, $a$. 
This quantity is determined entirely from the 2-body scattering problem, 
and is straightforward to calculate for any particular 
2-body interaction. Our new general result is:
\be g\approx 1-2an_0 + O(n_0^2).\label{gld}\ee
The scattering length, $a$, can be positive or negative 
depending on the details of the interactions, even though
they are always assumed to be purely repulsive.

Appendix B discusses determination of the value of $g$ 
for our tight-binding model. In Appendix C we present 
results on the correlation function of the boson 
creation operator which is useful in confirming 
the RG picture regarding the relevance or irrelevance 
of a defect.  Appendix D discusses the difference 
in ground state energy for periodic versus anti-periodic 
boundary conditions, another useful diagnostic for 
relevance or irrelevance of a defect. Appendix E 
contains some estimates of the effects of point disorder. 
Appendix F points out the connection between our model 
and one which has recently attracted attention 
from the string theory community.

A brief summary  of these results appeared earlier.\cite{Hofstetter} 

\section{Classical and quantum formulations of 
interacting flux lines in (1+1) dimensions}

  Following, e.g., [\onlinecite{hwa93}] we consider a
model free energy $F$ for $N$ flux lines in an extreme type II
superconducting sample of thickness
$L_\tau$ in the $\tau$ direction in the presence of a single columnar
defect aligned in the $\tau$ direction and located at $x=0$:
\be
F = \int^{L_\tau}_0d\tau \left\{ \sum_{i=1}^N\left[ {{\tilde{\epsilon}_1} \over 2}   
\left [ {{dx_i(\tau)} \over {d\tau}}
 \right ]^2   -\epsilon_d \cdot \delta [x_i (\tau )]
-{{\phi_0 H_\perp} \over {4
\pi}} \cdot   {{dx_i (\tau )} \over {d\tau}}\right]
+ {1 \over 2} \sum_{i \neq j} 
 V_{\hbox{int}} [ | x_i (\tau ) - x_j (\tau )|] \right\}  \label{eq:first}
\ee
\noindent where $x_i (\tau )$ denotes the trajectory of the $i^{th}$
vortex line, $V_{\hbox{int}}(x )$ is the repulsive interaction potential between
flux lines, which can be taken to be local in $\tau$.  We don't 
expect this locality assumption to qualitatively change the long 
distance physics in the dilute limit. The coupling 
  $\epsilon_d$ measures  
the (attractive) interaction between a flux line and the  columnar
defect.  The tilt modulus, in the dilute limit, $n_0\lambda <<1$, where 
$n_0$ is the vortex density and $\lambda$ is the penetration depth, (i.e. $H\gtrsim H_{c1}$), 
for a planar sample which is invariant under rotations in the plane is
\be
\tilde \epsilon_1 = \left ( {{\phi_0} \over {4 \pi \lambda}}
\right )^2 \ln \: \kappa ,
\ee
\noindent where $\kappa  = \lambda / \xi$ is the ratio of the London
penetration depth $\lambda$ and the coherence length $\xi$.
The canonical partition function for a system of $N$ lines
is given by the Boltzmann integral:
\be
Z= {1\over N!}\Pi^N_{i=1} \int {\mathcal{D}} [x_j (z)] \: e^{- F
[ \{ x_i (z) \} ]/T},
\label{Z}\ee
where we have set $k_B=1$. 

From here it is possible to pass directly to a 
continuum elastic formulation of this model or else 
to use a quantum description where we regard $x_j(\tau )$ 
as the trajectory of a particle and the classical 
Boltzmann sum as a Feynman path integral.  We pursue 
the first direction in sub-section A and the second 
in sub-section B. Upon using a quantum hydrodynamics 
approximation to the quantum model we obtain the 
same continuum field theory. This field 
theory contains some parameters which we estimate, 
starting from the underlying vortex model,  
in sub-section C. 

\subsection{Classical continuum elastic theory}
It is very convenient to pass to a continuum approximation 
using a coarse-grained displacement field, $u(x,\tau )$, 
like that used to describe a 
two-dimensional smectic liquid crystal in an external field.\cite{Domb}
 A standard way of doing this 
is to write the trajectory of the $j^{\hbox{th}}$ vortex as:
\be x_j(\tau )=ja_0+u_j(\tau ),\label{defu1}\ee
(see Fig. 1a) where $a_0=n_0^{-1}$ is the average vortex 
spacing in the $x$-direction and define $u(x,\tau )$ 
at the ``equilibrium'' positions of the vortices by:
\be   u (ja_0,\tau )=u_j(\tau ).\ee
This identification implies that the coarse grained density is:
\be n(x,\tau )\approx n_0\left[ 1-{\partial u\over \partial x}
\right] .\label{CGD}\ee
While this definition of $u$ is standard also in higher dimensions, 
there is another way of defining it, special to 1 dimension 
which has certain advantages but is essentially equivalent. 
We first define a field $A(x,\tau )$ by the requirement 
that the position of the $j^{\hbox{th}}$ vortex
 along a constant $\tau$ slice 
is the point $x$ where $A(x,\tau )=j$.  Thus $A(x,\tau )$ can 
be taken to be a smooth monotonic function of $x$ varying from 
$0$ to $N$  where $N$ is the number of vortices. 
[An arbitrary smooth interpolation of $A$ can be 
chosen between the points where it is integer-valued.] 
It thus follows that the density is:
\be n(x,\tau )\equiv \sum_{j=1}^N\delta [x-x_j(\tau )]
={\partial \over \partial x}\sum_{j=-\infty}^\infty \theta [A(x,\tau )-j],
\label{densdef2}\ee
where $\theta (x)$ is the step function. ($\theta (x)=0$ for $x<0$ 
and $\theta (x)=1$ for $x>0$.)
Note that this definition implies that the number of vortex lines, 
along a constant $\tau$ slice, between $x_1$ and $x_2$ is:
\be \int_{x_1}^{x_2} n(x,\tau )dx = [A(x_2,\tau )]-[A(x_1,\tau )],\ee
where $[A]$ denotes the integer part. 
On the other hand, Eqs. (\ref{defu1}) and (\ref{CGD}) imply
  that this quantity is:
\be \int_{x_1}^{x_2} n(x,\tau )dx \approx n_0\{ (x_2-x_1)-[ u(x_1,\tau )
-u(x_2,\tau )] \} .\ee
Thus we write:
\be A(x,\tau ) = n_0 [x-u (x,\tau )] ,\ee
and see that these two definitions of $u(x,\tau )$ are 
equivalent in a coarse-grained limit.

Upon using the Poisson summation formula, the density in Eq. (\ref{densdef2})can be written:
\be n(x,\tau )={\partial A\over \partial x}\sum_{j=-\infty}^\infty \delta 
[A(x,\tau )-j]={\partial A\over \partial x}\sum_{m=-\infty}^\infty 
e^{2\pi imA(x,\tau )}.\ee
When reexpressed in terms of $u$ this identity becomes:
\be n(x,\tau )=n_0\left[ 1-{\partial u\over \partial x}
\right] \sum_{m=-\infty}^\infty 
e^{iG_mx}e^{-iG_mu(x,\tau )}.\label{densu}\ee
Here the reciprocal lattice vectors, $\{G_m\}$ are 
\be G_m=2\pi m/a_0,\ \  m=0,\pm 1 . \ldots \label{nu}\ee

We neglect the columnar pin for the moment and use the standard
continuum elastic energy for a set of tilted vortex lines in
(1+1)--dimensions, namely\cite{blatter94,fisher89a,hwa93}
\be
F_0 = \int dx d\tau \left[
 \frac{1}{2} c_{44} \left(\partial_\tau u\right)^2 
+\frac{1}{2} c_{11} \left(\partial_x u\right)^2 
- (n_0\phi_0 H_\perp / 4\pi )\left(\partial_\tau u\right)\right] 
\label{elastic_energy}
\ee
where  $c_{11}$ and $c_{44}$ are the compressional and
tilt moduli respectively.  
We see from Eqs. (\ref{eq:first}) and (\ref{defu1}) that:
\be c_{44}\approx n_0 \tilde \epsilon_1.\label{c_44}\ee
 Although the $T=0$ value of $c_{11}$ is determined by vortex interactions as
 \be c_{11}\approx n_0^2\int dy V_{\hbox{int}}(y),\label{c_11}\ee
thermal fluctuations and confinement entropy of interacting
vortex lines lead to a value,\cite{Coppersmith}
\be
c_{11}\propto T^2n_0^3/\tilde \epsilon_1, \label{c_112}\ee
in the limit $n_0\to 0$, where the dimensionless constant 
of proportionality is independent of the vortex interactions. 
The shear elastic constant $c_{66}$, necessary
to describe a triangular Abrikosov flux lattice in (2+1)--dimensions,
is absent.  Elastic moduli such as $c_{11}$ and $c_{44}$ 
can be nonlocal (i.e.,
wavevector dependent in Fourier-space\cite{blatter94}), but for the large distance
physics of interest to us here, we can take them to be constants, equal
to their values on scales much larger than the particle spacing or the
range of the interaction. 

Including the columnar pin simply adds a term to F:
\be F \to F-\epsilon_d\int d\tau n(0,\tau ).\label{F+n}\ee
We can include this in our elastic free energy by using
Eq. (\ref{densu}) to express $n(0,\tau )$. Many long wavelength properties of the vortex-pin system 
can be studied using this continuum free energy. However, for 
some purposes it is convenient to use the quantum mechanical 
formulation of the model, outlined in the next sub-section.

We note that the elastic continuum free energy of 
Eq. (\ref{elastic_energy}), with its adjustable constants $c_{\hbox{44}}$ 
and $c_{\hbox{11}}$, modified as appropriate to account for point 
and/or columnar defects,  is expected to  give a valid description 
of the long distance physics of the vortex lattice for essentially 
arbitrary vortex densities.  On the other hand,  
the simple free energy of Eq. (\ref{eq:first}) is 
only valid at low vortex densities, $n_0<\lambda^{-1}$, corresponding 
to fields not to far above $H_{c1}$. At higher fields 
nonlocal couplings are required to capture the physics at all length scales.
 Consequently, the estimate of 
$c_{44}$ in Eq. (\ref{c_44}) is only expected to be valid at low densities.

\subsection{Quantum formulation}
Let us return to the original discrete formulation of our 
free energy, Eq. (\ref{eq:first}).  We may regard $x_j(\tau )$, 
$j=1,2,3,\ldots N$ 
as the trajectories of $N$ particles.  To describe 
a physical sample containing vortices, the Boltzmann 
sum in Eq. (\ref{Z}) could be done by first holding the entry and 
exit points of the $N$ vortices at the top and bottom of 
the sample fixed. The Boltzmann sum includes summing over all 
permutations of which vortex $i$, enters and exits at 
each of the prescribed entry and exit points.  The 
$1/N!$ factor is then neccesary to avoid over-counting. 
This expression can readily be seen to be the Feynman 
path integral for a density  matrix of a system of 
$N$ {\it bosons}.\cite{nelson88}
 Eq. (\ref{Z})
 can be rewritten in terms of the imaginary-time
evolution operator $e^{-L_\tau {\hat{H}} / \hbar}$ as
\be
Z= \left < \psi^f | e^{-L_\tau {\hat{H}} / \hbar} | \psi^i
 \right >,
\ee
where the bra and ket vectors are the initial and final
states, respectively, obtained by summing over all entry and exit points.
 The quantum Hamiltonian $\hat H$ describes an ensemble of 
interacting \emph{bosons} and is given by 
\begin{equation}
\hat H=-{\hbar^2 \over 2m}\int dx\psi^\dagger \left({d\over dx}-h\right)^2\psi
+{1\over 2}\int dx\, dy\, \hat{n}(x)\, V(|x-y|)\, \hat{n}(y)-\epsilon_0\hat n(0).
\label{Hamcon}\end{equation}
where 
$\psi (x)$ is the bosonic annihilation operator and  
\begin{equation}
\hat n(x)\equiv \psi^\dagger (x)\psi (x),
\end{equation}
is the boson number density.
To account for tilting of the external field away from the direction of the pin,
an imaginary vector potential $ih$ has been included, thus making the Hamiltonian 
{\it non-Hermitian}. In the following we will set $\hbar =1$.  We 
 measure time, $\tau$, as well as position, $x$, in units of length.  Thus various 
parameters in the quantum model have unusual dimensions.  
\begin{table}
\caption{Correspondence between quantities in classical vortex 
line problem and quantum boson problem. $\mu$ is the boson chemical potential. $\hbar$ 
is Planck's constant but $h$ is the transverse field. Although the dimensions 
of corresponding quantites do not match, they do upon forming physically relevant 
combinations such as $\hbar^2/2m \leftrightarrow T^2/2\tilde \epsilon_1$.}
\begin{tabular}{l|l}
Vortex Lines & Bosons\\
\hline 
$\tilde \epsilon_1$ & $m$\\
$V_{\hbox{int}}(x)$&V(x)\\
$\epsilon_d$ & $\epsilon_0$\\
$L_\tau$ & $\beta \hbar$\\
$H_{||}w / {\phi_0}$& $n_0$\\
$\phi_0 H_{||}/ (4 \pi )- \tilde \epsilon_1$& $\mu$\\
$\phi_0 H_\perp /(4 \pi )$& $h$\\

$T$\ \    & $\hbar$
\end{tabular}
\label{table:corr}
\end{table}
In Table (\ref{table:corr}) we show 
the correspondences 
between various physical quantities in the classical 
vortex and quantum boson models. Here 
 $n_0=a_0^{-1}$ is the average number of bosons per unit length 
in the one-dimensional (1D) quantum system and $w$ is the 
thickness of the slab. 

For  numerical simulations it is convenient to define a lattice 
regularization of the model (\ref{Hamcon}): 
\begin{equation}
\hat H=\sum_{i=0}^{L-1}\left[-t\left( b^\dagger_ib_{i+1}e^{-h}+b^\dagger_{i+1}b_ie^{h}\right)
+{U\over 2}\hat n_i(\hat n_i-1)+V\hat n_i\hat n_{i+1}-\epsilon_0\hat 
n_0\right] .
\label{Hamlat}\end{equation}
although this lattice has no physical meaning. 
Here $\hat n_i\equiv b^\dagger_ib_i$ denotes the boson density on lattice site $i$.
The equivalence between (\ref{Hamcon}) and (\ref{Hamlat}) holds 
for low densities per lattice site $n_0 << 1$.  
For numerical convenience, the Hilbert 
space is restricted so that there can only be 0, 1 or 2 bosons on 
each site. We normally impose periodic boundary conditions:
\be b_L\equiv b_0.\label{bc}\ee
In the low density limit, small $h$ limit, Eq. (\ref{Hamlat}) 
reduces to the continuum model (\ref{Hamcon}) with:
\begin{equation}t=1/2m.\end{equation}

To obtain a quantitative understanding of the analytical approximations used in 
this paper, and to determine numerically 
the Luttinger liquid parameter $g$, we have applied a non-hermitian 
generalization of the density matrix renormalization group (DMRG) 
algorithm to the tight-binding Hamiltonian (\ref{Hamlat}). 
The DMRG method\cite{White92} was originally invented to determine 
in a quasi-exact fashion numerically the properties (order parameters, 
correlations, structure functions) of the ground or low lying excited states of 
one-dimensional, strongly correlated quantum Hamiltonians,  preferably with  
short-ranged interactions. In contrast 
to many other techniques, DMRG performance 
is typically enhanced by strong interactions. System sizes that can be 
studied for Heisenberg or Hubbard-type models reach up to 
the order of a thousand sites. A good introduction and overview of many of the 
original DMRG applications may be found in Ref.\ (\onlinecite{Peschel99}). 

The key idea of DMRG is to grow iteratively, starting from some very small system 
size which can still be diagonalized exactly, a sequence of systems of 
linearly increasing size while carrying out reduced basis 
transformations at each growth step to keep the size of the 
underlying Hilbert space fixed. This reduced basis transformation is 
chosen such as to introduce the minimal error in the representation 
of the physical state of interest, most often the ground state. This is 
achieved by determining this state $|\psi\rangle$ by some large sparse matrix 
diagonalization algorithm, and partitioning the entire system into two 
blocks $A$, $B$ for which density matrices are derived by tracing out the 
states of the other block in the pure state projector,
$\hat{\rho}_{A(B)}={\rm Tr}_{B(A)}|\psi\rangle\langle\psi|$. The 
eigenvalue spectra of the density matrices determine the new 
reduced bases for the system parts by choosing as new bases for the blocks a 
fixed number $m$ of eigenstates of the density matrices characterized by the 
largest eigenvalues.

This fundamental idea of the DMRG was generalized to renormalize not just quantum 
Hamiltonians, but also classical transfer matrices for statistical 
mechanics problems in two dimensions\cite{Nishino95} and quantum transfer 
matrices to study the thermodynamic properties of one-dimensional quantum 
systems\cite{Bursill96}, complementing the $T=0$ results of the original method.
Carlon, Henkel and Schollw\"{o}ck\cite{Carlon99} have generalized the DMRG 
method to renormalize strongly non-hermitian transition matrices that originate 
in a master equation 
 formulation of reaction-diffusion problems. The steady-state behaviour of 
these systems can then be derived from the left and right eigenstates 
corresponding to the eigenvalue with the 
smallest real part which are thus the ``ground state'' pair of that problem. 
This DMRG variant can be directly applied to 
determining the ground state eigenfunction pair of a non-hermitian quantum mechanics 
problem.

The main numerical problem in the generalization of the DMRG to 
non-hermitian systems is given by the observation that non-hermitian 
diagonalization of large sparse matrices is much less stable than in the 
hermitian case. While eigenvalues can be obtained with satisfactory 
precision, eigenstates show small numerical inaccuracies that tend 
to accumulate during DMRG runs as these eigenstates are at the basis 
of the reduced basis transformations carried out in each DMRG step. 
In the unsymmetric Lanczos algorithm we have used\cite{Golub96}, the origin of 
these inaccuracies is mainly due to the inevitable loss of the global 
biorthonormality of the sets of Lanczos ansatz states for right and left 
eigenstates. We have achieved good numerical stability by applying a 
selective, very time-efficient re(bi)orthogonalization procedure introduced by 
Day\cite{Day97}.

In determining the density matrices for the non-hermitian DMRG, there is 
a further arbitrariness: there are both symmetric and nonsymmetric 
density matrices, defined by partial traces, 
$\hat{\rho}_{symm}=(1/2) {\rm Tr} (|\psi_{R}\rangle\langle \psi_{R}|+|\psi_{L}\rangle\langle \psi_{L}|)$
and $\hat{\rho}_{nonsymm}={\rm Tr} |\psi_{R}\rangle\langle \psi_{L}|$ respectively,
where $\langle \psi_{L}|$ and $|\psi_{R}\rangle$ are the left and 
right eigenstates of interest for the total system. The reduced basis 
transformations derived from the two choices are not identical. There is no 
clear-cut preference in non-hermitian DMRG: In the case of 
the quantum transfer matrix DMRG, which also suffers from (weak) 
non-hermiticity, the unsymmetric choice was found to be 
superior in precision\cite{Nishino99}, whereas in another non-hermitian DMRG 
version, the stochastic transfer matrix DMRG, 
only the 
symmetric choice yields useful information\cite{Kemper01,Enss01}. 
Studies we have carried out for our problem of interest indicate that 
both approaches are feasible for high accuracy, but that numerical stability 
concerns favor the symmetric choice.  

With these choices made, we have studied system sizes up to $L=256$ sites with 
up to $N=64$ bosons, the particle density being at or below a quarter.
Up to $m=300$ states have been kept in the reduced Hilbert spaces and found to 
give effectively converged results 
for currents, energies, and local quantities 
such as particle densities. The low particle density, leading to a strong 
arbitrariness in the insertion of particles during system growth, and the presence of an impurity 
mandate the application of the so-called finite-size DMRG 
algorithm,\cite{White92}  
which has been applied up to 11 times to achieve converged results.

In applying DMRG to bosonic systems, the possibly divergent number of 
bosons per 
site has to be truncated algorithmically to some maximum number. 
As we are considering superconductors in the low flux line density limit, 
we fixed the maximum number of bosons per site to be 2. 
This constraint is consistent with 
average particle densities of no more than $0.25$; the validity of this 
truncation has been checked for selected parameter sets by increasing 
the maximum number of bosons per site. It should be mentioned that 
these findings are not at variance with the statement that in Luttinger 
liquids with $g < 1$ impurities correspond to relevant perturbations and 
scale to infinity in effective field theories under 
renormalization group (RG) flow.\cite{Kane,Eggert} In the underlying 
lattice model, the resulting perfect pinning is effected by the 
generation of a very long-ranged effective local pinning potential whose 
strength decays only as a power law away from the impurity, but whose scale is 
essentially that of the original impurity.\cite{Meden02} Hence, 
we do not expect a 
 particularly strong enhancement of the local boson density at the 
impurity site, as confirmed by our numerical results; 
in all runs, even at the impurity site, the boson density is well 
below 1 for all impurity strengths considered in this paper.

We now pass to a quantum hydrodynamic formulation of this 
model.\cite{Haldane}  We may express the boson density operator 
in terms of a quantum field, $u(x)$, using Eq. (\ref{densu}). 
In order to conform to more standard notation in the quantum 
literature, we replace the displacement field, $u(x,\tau )$,
which has dimensions of length, with a dimensionless field, $\theta (x,\tau )$:
\be \theta (x,\tau )\equiv -n_0u(x,\tau ).\label{thetatou}\ee
Thus Eq. (\ref{densu}) for the density operator becomes:
\begin{equation}
\hat n(x)=\left( n_0+
{d\theta \over dx}\right) \sum_{m=-\infty}^\infty e^{2\pi im[ n_0x+\theta (x)]}.
\label{n-theta}
\end{equation}

We 
write the boson creation operator in the form:
\begin{equation}
\psi^\dagger (x)\propto \sqrt{n_0+{d\theta \over dx}}
\sum_{m=-\infty}^\infty e^{2\pi im[ n_0x+\theta (x)]}e^{i\phi (x)},
\label{bospsi}\end{equation}
with, of course, the Hermitian conjugate expression for the 
operator, $\psi (x)$. Note that this formula is 
consistent with $\hat n=
\psi^\dagger \psi$ using the fact that 
$\delta^2(x)\approx \Lambda  \delta (x)$ where $\Lambda$ 
is an ultra-violet cut-off with dimensions of wave-vector.  We have introduced 
another bosonic field, $\phi (x,t)$, which represents the phase of $\psi^\dagger$. 
We generally keep only the most relevant terms (in a renormalization group sense) in these expressions, 
writing:
\begin{eqnarray}
\hat n(x)&\approx &n_0+{d\theta \over dx}+\hbox{constant}\times 
\cos \{ 2\pi [n_0x+\theta(x)]\}\nonumber \\
\psi^\dagger (x)&\propto &\hbox{constant}\times e^{i\phi (x)}.
\label{bosnpsi}\end{eqnarray}
The $\theta$ and $\phi$ fields do not commute. In fact, 
in order to correctly reproduce the continuum commutation relations between 
$\hat n(x)$ and $\psi^\dagger (x)$, namely
\begin{equation}
[\hat n(x),\psi^\dagger (y)]=\delta (x-y)\psi^\dagger (x),
\end{equation}
we require:
\begin{equation}
[d\theta (x)/dx ,\phi (y)]=-i\delta (x-y).\label{comm}\end{equation}
Thus, 
we can identify:
\begin{equation}
d\theta /dx=\hat \Pi,\label{Pi}\end{equation}
  as the momentum operator conjugate to $\phi$. 
Upon integrating Eq. (\ref{comm}) we obtain:
\begin{equation}
[\phi (x), \theta (y)]=(- i/2) \ \hbox{sgn} (x-y),\end{equation}
where sgn$(x)$ is the sign function, $\hbox{sgn}(x)=x/|x|$. Thus we see 
that $[d\phi /dx,\theta (y)]= -i\delta (x-y)$ and hence 
the conjugate momentum to $\theta $ is 
\begin{equation}
\hat \Pi_\theta =d\phi /dx .\label{Piu}\end{equation}
We may now write a long wavelength 
low energy approximation to the Hamiltonian of Eq. (\ref{Hamcon}) 
(ignoring, for the moment, the tilt field and pinning 
potential) in terms of these phonon and phase variables.  Keeping 
only terms of quadratic order in the fields and their derivatives, 
the Hamiltonian density less the chemical potential times the particle density 
 may be written, for convenience, in terms of 
two new parameters, a phonon velocity $c$ and a dimensionless
``Luttinger liquid parameter'' $g$:
\begin{equation}
\hat {\cal H}_{\phi}(x)-\mu \hat n(x)=
{c\over 2}\left[ {g\over \pi } \left({d\phi \over dx}\right)^2+
{\pi \over g} \hat \Pi^2\right] .\label{H-phi}\end{equation}
The first term comes from the kinetic energy in Eq. (\ref{Hamcon}) implying
 that
\begin{equation}
{cg\over \pi}={n_0\over m}.\label{cg}\end{equation}
The second term comes from the interaction term, which leads to:
\begin{equation}
{c\pi \over g}\approx \int dx V(|x|).\label{c/g}\end{equation}

Upon canonically transforming to the Lagrangian and 
then going to imaginary time: $\tau \equiv it$, we have:
\begin{equation}
{\partial \phi \over \partial t}={\pi c\over g}\hat \Pi,\end{equation}
and the imaginary time action
\be S_\phi = \int_0^Ldx \int d\tau {\cal L}_\phi (x,\tau ),\ee
where the Lagrangian density is
\begin{equation}
{\cal L}_\phi (x,\tau )= {g\over 2\pi } 
\left[{1\over c}\left({\partial \phi (x,\tau ) \over \partial
    \tau}\right)^2
+c\left({\partial \phi (x,\tau )\over \partial x}\right)^2\right] .
\label{L-phi}\end{equation} 

Alternatively, we may write $\hat H$ in terms of the phonon field, $\theta $. 
With the help of Eqs. (\ref{Pi}) and (\ref{Piu}) we see that the Hamiltonian 
density can also be written as:
\begin{equation}
\hat {\cal H}_{\theta}
-\mu \hat n={c\over 2}\left[{g\over \pi }\hat \Pi_\theta ^2+ 
{\pi \over g} \left({d\theta \over dx}\right)^2\right] .
\label{H-u}\end{equation}
Another canonical transformation  gives:
\begin{equation}
{\partial \theta \over \partial
  t}={gc \over \pi}\hat \Pi_\theta  \label{canu}
\end{equation}
and hence the action in terms of the $\theta $-field reads:
\be S_\theta  = \int_0^Ldx \int d\tau {\cal L}_\theta (x,\tau ),\ee
with
\begin{equation}
{\cal L}_\theta (x,\tau )={\pi \over2g}\left[{1\over c}\left( 
{\partial \theta (x,\tau )\over \partial \tau}\right)^2
+c\left( {\partial \theta (x,\tau )\over \partial x}\right)^2\right].
\label{L-u}\end{equation}
Eq. (\ref{L-u}) is, of course, just the result obtained from classical 
continuum elastic theory, Eq. (\ref{elastic_energy}), with:
\bea g&=&{\pi Tn_0^2\over \sqrt{c_{11}c_{44}}}\nonumber \\
c&=&\sqrt{c_{11}\over c_{44}}.\label{g}\eea

It is straightforward to add the imaginary vector potential, $h$, 
in the quantum hydrodynamic formulation of the model.  We begin 
with the Hamiltonian of Eq. (\ref{Hamcon}) and then rewrite 
the boson creation operator, $\psi^\dagger$ using Eq. (\ref{bosnpsi}).
The modification of the Hamiltonian density in $\phi$-representation is:
\be {\cal H}_{\phi}\to {\cal H}_{\phi}-{ihn_0\over m}{d\phi \over dx}
-{h^2\over 2m}n_0.\label{Hphih}\ee
We may alternatively calculate the extra term in the 
Lagrangian density in $\theta$ representation, using Eq. (\ref{Piu}) 
and then canonically transformating from Hamiltonian to Lagrangian, 
giving: 
\be {\cal L}_{\theta} \to {\cal L}_{\theta} +h{\partial \theta \over 
\partial \tau}.\label{Lthetah}\ee
This is, of course, consistent with Eq. (\ref{elastic_energy}), using 
Table I and Eq. (\ref{thetatou}).

\subsection{The Luttinger liquid parameter $g$:\ \ its significance and numerical value}
The quantity $g$ is an essential parameter which determines all critical properties 
of the model, with or without a pin or a transverse field. In this 
sub-section we indicate the 
significance of $g$ and estimate its value for the physical vortex problem 
in the limits of dense and dilute vortices. 

From Eq. (\ref{L-u}) we see that the density field, $\theta$, has 
the correlator:
\be <\theta (\tau ,x)\theta (0,0)>\approx -{g\over 4\pi^2} \ln [c^2\tau^2+x^2]
+\hbox{constant} \label{thetacorr}\ee
With the help of Eq. (\ref{bosnpsi}), we see that the 
density correlation function is given by:
\begin{equation}
<n(x,\tau )n(0,0)>\to n_0^2+{gc^2\over 2\pi^2}{c^2\tau^2-x^2\over 
(c^2\tau^2+x^2)^2}
+\sum_{m=1}^\infty 
 {A_m\cos (2\pi n_0mx)\over [x^2+c^2\tau^2]^{gm^2}}+\ldots ,
\label{etadef}\end{equation}
where the $A_m$'s are constants. 
Note that a conventional vortex lattice cannot form in our two-dimensional system, 
but we get instead quasi-long range vortex order with density 
oscillations  spaced by the average inter-vortex separation, $a_0$. The Fourier transformed 
density-density correlation function, or structure function,
$S(q_x,q_\tau )$, has algebraic singularities at the 
reciprocal lattice wave-vectors, $G_m\equiv 2\pi m n_0$:
\be S(q_x,q_\tau )\propto 
\frac{1}{\left[c^2 (q_x - G_m)^2 + q_\tau^2\right]^{1-g m^2}}
\ \  [\hbox{for}\ (q_x,q_\tau )\approx (G_m,0)]\label{Sq_decay}
\label{S}\ee
We see that $S(\vec q)$ diverges at $q_x=\pm G_1$, $q_\tau =0$ 
whenever $g<1$. See Fig. (\ref{fig:S}). 

We can use the results above to assess the effect of a single columnar defect
on the vortex density within perturbation theory.
To lowest order in the pinning strength $\epsilon_d$ we find
\bea
\delta n(x,\tau) &=& <n(x,\tau)> - n_0 \nonumber \\
&=& \epsilon_d \int d\tau'  C(x, \tau-\tau'),  
\label{delta_n_first}
\eea
where
\be
C(x-x', \tau-\tau') = <n(x,\tau) n(x',\tau')>_0 - 
<n(x,\tau)>_0 <n(x',\tau')>_0. 
\label{n_correlation}
\ee
Here, $<\ >_0$ represents an average in an ensemble where both tilt 
and the defect are absent.  When
Eq.~(\ref{delta_n_first})
 is rewritten in Fourier space, the linear response to the
perturbation induced by the columnar pin is determined by the
structure function.  For a system with spatial extent $L_\tau$ 
in the time direction, we have
\be
\delta n(q_x, q_\tau) = \epsilon_0
S(q_x, 0) \, L_\tau \delta_{q_\tau,0}  
\label{delta_n}
\ee
The most singular response is at wave-vector $(q_x,q_\tau )=(\pm G_1,0)$, i.e. for 
$m=1$ in Eq. (\ref{S}).
Thus the response diverges for $g<1$ suggesting that a single 
pin is a relevant perturbation for $g<1$ but is irrelevant for 
$g>1$.  This conclusion will be modified by a transverse field 
or by point disorder as we will see in later sections. 
The leading perturbative result for 
the oscillating part of $\delta n(x)$ at long distances
from the pin in real space is thus:
\be \delta n(x,\tau ) \propto {\epsilon_0\cos (2\pi n_0x)\over |x|^{2g-1}}.
\label{dnPS}\ee

\begin{figure}
\begin{center}
\includegraphics[width=0.4\linewidth]{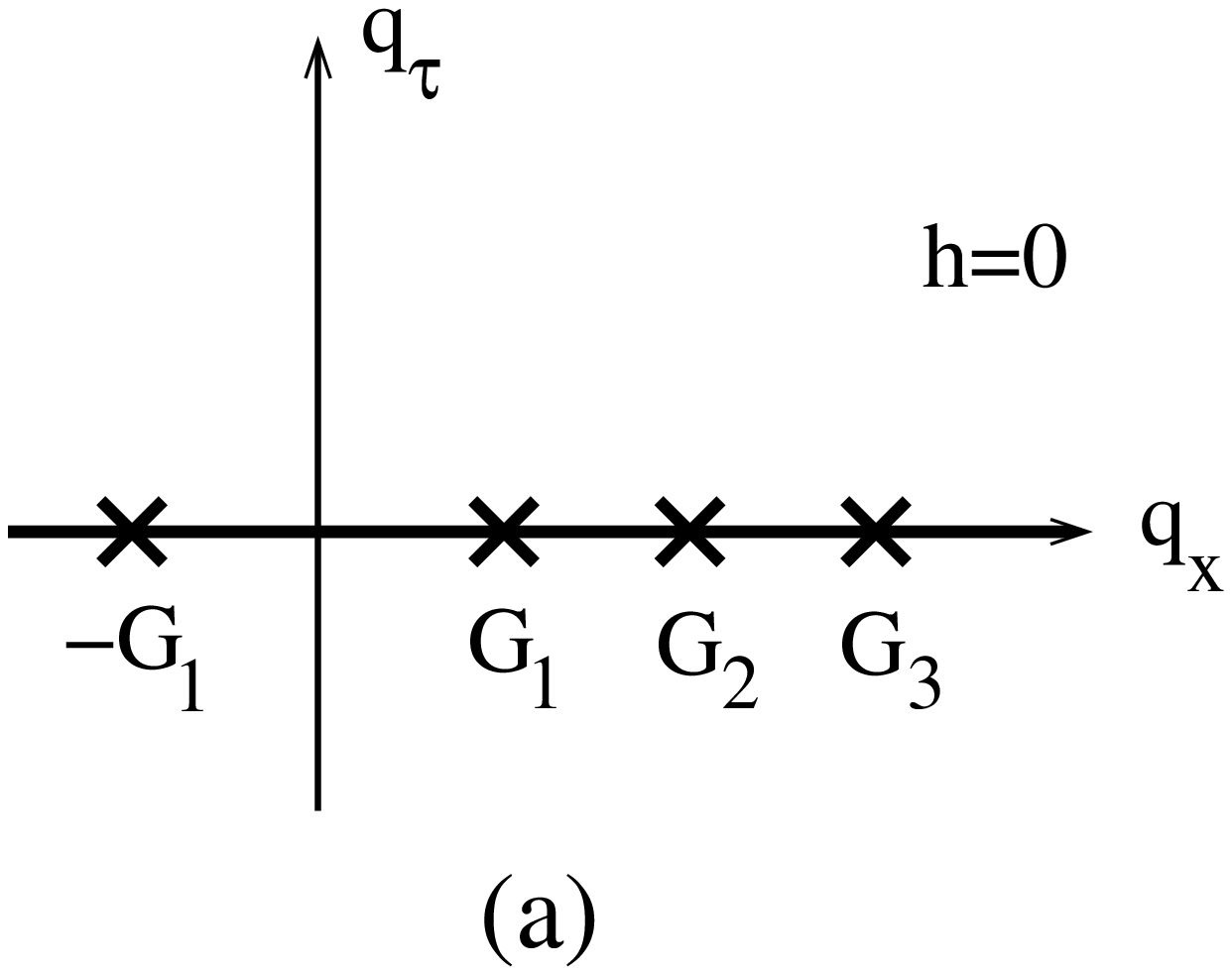}
\hspace{0.1\linewidth}
\includegraphics[width=0.4\linewidth]{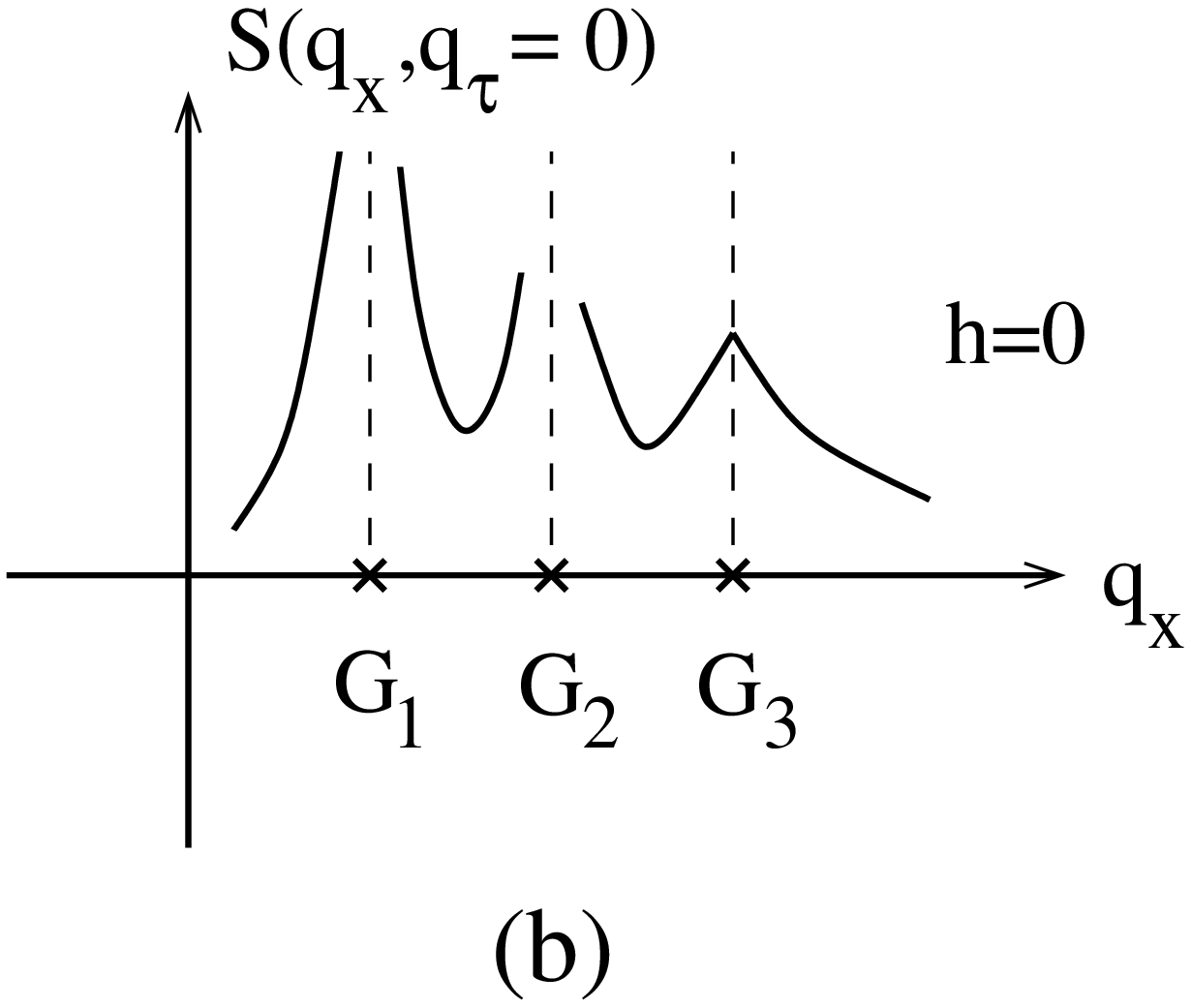}
\caption{(a) Reciprocal lattice vectors $(G_m, 0)$ in the $(q_x, q_\tau)$--plane
for $h = 0$.  The structure function describing vortex correlations in
reciprocal space is singular at these points .  The Fourier--transformed
potential of a single attractive columnar defect is only nonzero on the
heavy line along the $q_x$--axis. (b)  Profile of the structure function
$S(q_x,q_\tau)$ along the $q_x$ axis.  The structure function diverges near
$(G_1, 0)$ for $g < 1$.  It is also singular (although it need not diverge) at
the higher order reciprocal lattice vectors.}
\label{fig:S}
\end{center}
\end{figure}

Similarly, from the second of Eqs. (\ref{bosnpsi}) and 
 Eq. (\ref{L-phi}) we obtain 
the correlation function of the boson creation operator:
\begin{equation}
<\psi^\dagger (x,\tau )\psi (0,0)>\to {\hbox{constant}\over [x^2
+(c\tau )^2]^{1/4g}}.\label{psicorr}
\end{equation}
This correlation function 
 can be also derived directly from the classical continuum 
elastic theory\cite{schulz}
 where it is proportional to $\exp [-V(x,\tau )/T]$, 
$V(x,\tau )$ being the extra free energy arising from 
a dislocation pair located at $(x,\tau )$ and $(0,0)$. 
Such a configuration is shown in Fig. (\ref{fig:top_defect}).
[A similar method can also be used to explore 
correlations of the boson order parameter associated 
with flux lines in (2+1) dimemsions.\cite{nelson91}] 
We do not usually allow such 
topological defects in our Boltzmann sums over 
vortex configurations.  Nonetheless, we shall see in Sec. IV that 
this correlation function is 
very useful in studying the limit of a strong pin. 
\begin{figure}
\begin{center}
\includegraphics[width=0.45\linewidth]{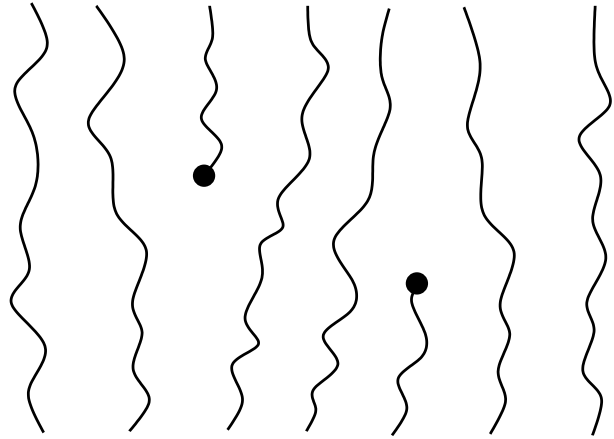}
\caption{topological defect configuration which occurs in 
the XY model but not in our vortex model.  The extra 
free energy associated with this topological defect pair
determines the correlation function of the boson 
creation operator in Eq. (\ref{psicorr}).}
\label{fig:top_defect}
\end{center}
\end{figure}

We note, in passing, that the effect of the discrete mesh used in
simulations of lattice models such as Eq.~(\ref{Hamlat}) can also be treated in
linear response theory about a continuum model.  Indeed, if the
average vortex separation is an integer p times the lattice spacing of
the mesh (most of our numerical calculations are carried out for 
$p = 4$), we can take
\be
V_D(x) \sim -\epsilon_p \cos(2\pi p x/a_0)  
\label{bulk_perturbation}
\ee
to describe the periodic mesh to leading order in perturbation theory.
Eq.~(\ref{bulk_perturbation}) also describes the first nontrivial Fourier coefficient
arising from a \emph{periodic} array of columnar pins.  The Fourier transform
$\hat{V}_D(q_x)$ on the right-hand of Eq.~(\ref{delta_n}) 
is now nonzero only for $(q_x,q_\tau) = (2\pi p/a_0, 0) = (G_p,0)$, 
where the structure function diverges according to
\be
S(q_x,q_\tau) \sim \frac{1}{\left[(q_x - G_p)^2 + q_\tau^2\right]^{1-\eta_p/2}}
\ee
with $\eta_p = 2 g p^2$. For bulk perturbations like Eq.~(\ref{bulk_perturbation}), 
it is well known that the corresponding renormalization 
group recursion relation reads:\cite{jose}
\be
\frac{d\epsilon_p(l)}{dl} = \left(2 - \eta_p/2\right) \epsilon_p(l). 
\ee
Hence, we conclude that the mesh is irrelevant relative to a continuum model whenever 
\be
g > g_c = 2/p^2.
\ee
For $p = 4$, corresponding to $1/4$--filling of the mesh with flux lines, $g_c=1/8$.
Because all simulations at $n_0=1/4$ in this paper lead to values of $g$
significantly larger than $1/8$, we can safely neglect the effect of the
lattice on the large distance physics.

We now turn to estimating the value of $g$. 
 When the vortex lines are
dense enough so that $n_0\lambda \gtrsim 1$, the magnetic field is approximately uniform in
the superconducting slab.  On scales large compared to $a_0$ and $\lambda$, we then
expect that the energy can be approximated by the usual form from
magnetostatics 
\be
F_0 \approx \int d^3r \left[\frac{\mathbf{B}^2(\mathbf{r})}{8\pi} - 
\frac{\mathbf{H}}{4\pi} \cdot \mathbf{B}(\mathbf{r}) \right], 
\label{em_field}
\ee
where $\mathbf{B}(\mathbf{r})$ is the magnetic field intensity, $\mathbf{H}$ is the
applied field and 
the integral runs over $x$, $\tau$ and the short direction $\hat{y}$ 
of a thin slab of thickness $w\lesssim \lambda$.  We take $\mathbf{H}$  
parallel to the time-like direction $\tau$ and note that Eq.~(\ref{em_field}) is
minimized for 
$\mathbf{B}(\mathbf{r}) = H \mbox{\boldmath $\tau$} \equiv B_0 \mbox{\boldmath $\tau$}$.  
We then neglect the variation of $\mathbf{B}(\mathbf{r})$ across the slab 
along $\hat{y}$ and expand about the state of a uniform field by setting 
$\mathbf{B}(\mathbf{r}) = (B_x, 0, B_0 + \delta B_\tau)$, with
$|B_x|,|\delta B_\tau |\ll B_0$.  
Upon making the identifications $\delta B_\tau(x,\tau) / B_0 = -\partial_x u(x,\tau)$ 
and $B_x(x,\tau) / B_0 = \partial_\tau u(x,\tau)$, we obtain an expression like
Eq.~(\ref{elastic_energy}) with 
\be
c_{44} = \frac{w B_0^2}{4\pi}, \ \ \ c_{11} = \frac{w B_0^2}{4\pi}\label{hin}
\ee
After setting  $B_0 \approx \phi_0 n_0/w$, 
where $\phi_0 = 2\pi \hbar c/2e \approx 2\times 10^{-7} \hbox{gauss-cm}^2$
is the flux quantum, 
we find from Eq.~(\ref{g}) that 
\be
g \approx \frac{4\pi^2 T w}{\phi_0^2}
\ee
With $w\approx 10^{-4}cm$  (we have in mind thin slabs with thickness 
of order the London penetration depth), and $T\approx 77K$ 
we find that $g\approx 10^{-3} \ll 1$ in this dense limit.  

In the dilute limit, $n_0\to 0$, $g\to 1$, corresponding 
to free fermions.\cite{deGennes,Pokrovskii,Coppersmith,Haldane}
That $g=1$ corresponds to free fermions can be simply checked
from the fact that the density-density correlation function in 
Eq. (\ref{etadef}) 
reduces to that of free fermions in this case:
\begin{equation}
<n(x,\tau )n(0,0)>\to n_0^2+\left[{c^2\over 2\pi^2}
+A_2\cos (2\pi n_0x)\right]{c^2\tau^2-x^2\over 
(c^2\tau^2+x^2)^2}
+\ldots .
\end{equation}
 Note that this asymptotic behavior of g is consistent with
the behavior of $c_{44}$ and $c_{11}$ in 
Eq. (\ref{c_44}) and (\ref{c_112}).  Free fermion
behavior in the dilute limit follows from  the equivalence of hard-core bosons 
with free fermions in (1+1) dimensions. As long 
as the average boson separation is large compared to 
the range of the inter-boson interaction, this 
``hard-core'' result holds at long distances even for 
a finite range inter-vortex interaction.   
$g=1$ is the border-line case where the pin is marginal. 
 We will 
exploit this equivalence of dilute bosons to non-interacting 
fermions in the next section.

In Appendix A we 
derive the leading correction to this result at low density, 
Eq. (\ref{gld}).  We expect this result to be  exact 
for our vortex system despite the fact that we ignored 
vortex-vortex interactions that were non-local in $\tau$ 
in our approximate free energy of Eq. (\ref{eq:first}). Such non-local interactions 
ultimately have similar effects to quartic and higher 
terms in the dispersion relation (see Appendix B of 
the second article in [\onlinecite{nelson92}]) and 
we expect that they only affect $g$ at
$O(n_0^2)$.

Whether $g$ increases or decreases as 
the density is increased from 0 depends on the sign 
of the scattering length, $a$, for the vortex-vortex interaction 
potential $V(x)$. This sign depends on the detailed form 
of $V(x)$. [Note, for example, that an infinite hard core 
repulsion gives $a>0$ whereas a repulsive $\delta$-function 
potential gives $a<0$.]  The appropriate potential, $V(x)$, 
is determined not only by the bulk inter-vortex interactions 
but also by the thin-slab geometry. Results on this 
scattering length will be reported elsewhere.\cite{Affleck}
If $a>0$ then $g$ may decrease monotonically with 
increasing $n_0$ so that the pin will be relevant for 
any $n_0$. On the other hand, if $a<0$, then $g$ 
initially {\it increases} with $n_0$.  Since we 
have shown that it goes to a very small value 
at large densities it must then exhibit non-monotonic 
behavior.  Furthermore, it must pass through the 
value $g=1$ at some finite critical density, $n_c$. In 
this case, the pin will be irrelevant for $n_0<n_c$ 
and relevant for $n_0>n_c$. 

The value of $g$ for our lattice model as a function 
of the microscopic parameters $U/t$, $V/t$ and $n_0$ 
can be determined at low $n_0$ from an exact formula for 
the scattering length, $a$, 
of the microscopic model or more
generally by numerical methods.  See Appendix B.

\section{Dilute Limit: Free Fermions}
In this section, we consider the dilute limit where we may 
approximate interacting bosons by non-interacting fermions. 
This regime corresponds to $g=1$: a marginal pin. 

\subsubsection{Density oscillations}
The density oscillations induced by the pin can be calculated 
straightforwardly using the standard method of calculating 
Friedel oscillations for non-interacting fermions, suitably 
generalized to the non-Hermitian case:
\be <0|\hat n(x)|0>={\sum_n}'\psi_n^L(x)\psi_n^R(x).\label{densLR}\ee
Here $\psi_n^L$ and $\psi_n^R$ are the left and right eigenfunctions 
of the non-Hermitian single-body Hamiltonian of Eq. (\ref{Hamcon}) 
with $V$ set to zero. The sum is restricted to the set of 
single particle levels with $Re (E_n)<E_F$, where $E_n$ are 
the single particle levels and $E_F$ is the Fermi energy. (We
denote this restricted sum by ${\sum}'$.)

The exact single particle 
energy eigenvalues and eigenfunctions for a $\delta$-function 
potential with an imaginary vector potential were obtained by 
Hatano and Nelson (H-N).\cite{hatano97}
  [See the Appendix of the second reference 
in (\onlinecite{hatano97}).] Here we adapt
these results to obtain the Friedel oscillations.  
Note the change of notation from that paper:
\bea \hbar &\to& 1\nonumber \\
g&\to & h\nonumber \\
V_0&\to& \epsilon_0\nonumber \\
L_x&\to& L.\eea
We must also keep in mind that, assuming periodic 
boundary conditions on the bosons,  the boundary conditions on the 
single-particle wave-functions in the effective fermion problem are 
periodic if the total number of bosons is odd but anti-periodic 
if the total number of bosons is even. [See Appendix A.]
 As H-N show, the $\delta$-function potential produces a single 
bound state if $m\epsilon_0>h$. There are also extended states 
which are responsible for the long distance density oscillations.

 We consider the non-Hermitian single-particle 
Schroedinger equation for the right eigenfunctions:
\be \left[-{1\over 2m}\left({d\over dx} -h\right)^2-\epsilon_0\delta (x)
\right] \psi^R(x)=E\psi^R(x).\ee
The extended right eigenfunctions are written
in terms of complex right-wave-vectors, $K_n$:
\be \psi^R_n(x)=A_ne^{(iK_n+h)x}+B_ne^{(-iK_n+h)x},\ \  (x\neq 0) \ee
for constants $A_n$ and $B_n$ determined by the boundary conditions. 
 Assuming $N$ odd,  so 
that the eigenfunctions obey periodic boundary conditions, the 
allowed values of $K$ satisfy:
\be K[\cosh (Lh)-\cos (LK)]+m\epsilon_0\sin LK=0,\label{K}\ee
where we may assume, without loss of generality that Im$K>0$.
Provided $h>0$ and $hL>>1$, Eq. (\ref{K}) can be approximated as:
\be K\left[ e^{Lh}-e^{-iLK}\right] +im\epsilon_0e^{-iLK}=0,\ee
or
\be (K-im\epsilon_0)e^{-iLK}=Ke^{Lh}.\label{Kdet}\ee
We see that Im$K\approx h$ and hence:
\be K_n\approx k_n+ih+{i\phi (K)\over L},
\label{K2}\ee
where $n$ runs over all integers and
 \bea 
k_n&\equiv& 2\pi n/L, \ \ (h>m\epsilon_0) \nonumber  \\
k_n&\equiv& 2\pi (n+1/2)/L,\ \  (h<m\epsilon_0)
\eea 
(We may assume $|\hbox{Im}\phi |<\pi$.)
Substituting in Eq. (\ref{Kdet}) implies:
\bea \phi (K_n)&=&\ln \left[{k_n+ih\over k_n+i(h-m\epsilon_0)}\right],
\ \  (h>m\epsilon_0)\nonumber \\
&=&\ln \left\{-\left[{k_n+ih\over k_n+i(h-m\epsilon_0)}\right]
\right\},
\ \  (h<m\epsilon_0).\label{phi}\eea
The corresponding energies are:
\be E=(1/2m)K^2.\ee
The exact right eigenfunctions, for periodic boundary conditions, are 
given in (A.13) of H-N.  Taking the large $L$ limit these become:
\bea \psi^R_k(x) &=& e^{ikx}+(\pm e^{\phi}-1)e^{-ikx+2hx}, \ \  x< 0\nonumber \\
&=& \pm e^\phi e^{ikx},\ \  x> 0.\label{psiR}\eea
 Here, the $+$ or $-$ minus sign apply to $h>m\epsilon_0$ and 
$h<m\epsilon_0$ respectively. 
For $k>0$ Eq. (\ref{psiR}) has the interpretation of a particle 
coming in from the left and being reflected and transmitted.  
Note that the reflected wave decays exponentially as $e^{2hx}$ ($x<0$)
 unlike in the normal, Hermitian, case. 
The corresponding left eigenfunction is given by:
\be \psi^L_h(x;k)=\psi^R_{-h}(x;k)^*.\ee
It follows from Eq. (\ref{K}) that $K_n(-h)^*=K_n(h)$.  
The left eigenfunctions, at large $L$, are given by:
\bea \psi^L_k(x)&=&\pm e^\phi e^{-ikx},\ \  x<0\nonumber \\
 &=& e^{-ikx}+(\pm e^{\phi}-1)e^{ikx-2hx},\ \  x>0 \label{psiL}\eea
For $k>0$, Eq. (\ref{psiL}) describes a particle 
arriving from the right and being transmitted or reflected, with 
the reflected wave decaying exponentially. The 
product of these eigenfunctions, at large $L$, (for either sign of $x$) is given by:
\be \psi^L_k(x)\psi^R_k(x)=\pm e^{\phi}[1+(\pm e^{\phi}-1)e^{2(ik-h)|x|}].\ee
Upon normalizing these wave-fuctions and then 
integrating over all levels below the Fermi surface, we find the density at 
large $|x|k_F$:
\bea n(x)&=&{k_F\over \pi}+e^{-2h|x|}\int_{-k_F}^{k_F} {dk\over 2\pi}
(\pm e^{\phi (k)}-1)e^{2ik|x|}
\approx {k_F\over \pi}+e^{-2h|x|}
\left[
{\left(\pm e^{\phi (k_F)}-1\right) \over 4\pi i|x|}e^{2ik_F|x|}+c.c. \right]
\nonumber \\
&=&{k_F\over \pi}+e^{-2h|x|}
\left[{m\ep \over (h-i\pi n_0-m\ep )4\pi i|x|}e^{2ik_F|x|}+c.c.\right].
\eea
Note that the Friedel oscillations decay exponentially with 
a decay length given by $1/2h$.  This result arises 
  from the exponential decay of the reflected wave which leads to 
the exponential decay of the interference between 
incident and reflected waves. 

In the lattice model (\ref{Hamlat}) the limit $g=1$ corresponds to 
$U\to \infty, V=0$ 
and thus to a system of hardcore bosons, which in (1+1) dimensions can be mapped 
onto the Hamiltonian for non-interacting fermions (described by creation and 
annihilation operators $c^\dagger_i$ and $c_i$), namely
\be
H = -\epsilon_0 n_0 - t \sum_i \left(c^\dagger_{i} c_{i+1} e^{-h} + 
c^\dagger_{i+1}c_ie^{h}\right),
\label{free_fermions}  
\ee 
where $n_0\equiv c_0^\dagger c_0$ and we assume a canonical ensemble with $N$ particles. 
This Hamiltonian can easily be diagonalized exactly 
for system sizes of a few hundred lattice sites. 
Using Eq. (\ref{densLR}), we have 
obtained numerical results for the density profile shown in 
fig.~\ref{fig:friedel_free_fermion}. In agreement with the analytical 
predictions above we find strong Friedel oscillations which are 
suppressed drastically by a finite nonhermitian term $h>0$. 
We have also confirmed that the oscillations decay as 
$x^{-1} \exp(-x/\xi_{\perp})$ with a decay length $\xi_\perp \sim 1/h$.  

In fig.~\ref{fig:traffic_jam} we give a schematic representation 
of the underlying ``traffic jam'' picture. Since flux lines repell each 
other, they form a local vortex lattice,
 aligned with the imaginary time direction, in the vicinity of the pin. 
Finite tilt destroys this effect at large length scales, 
with a crossover scale given by $\xi_\perp$.  In 
  fig.~\ref{fig:traffic_jam} we have arbitrarily 
chosen to show a queue which is two deep on either 
side of this (symmetric) traffic jam, corresponding to a 
total of five maxima in a density plot such as Fig. 
(\ref{fig:friedel_free_fermion}).
\begin{figure}[h]
\includegraphics[width=9cm]{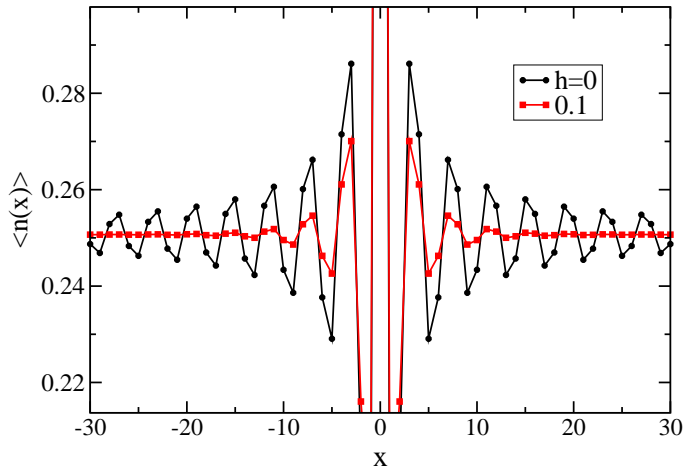}
        \caption{\label{fig:friedel_free_fermion}
Friedel oscillations of the flux line density vs. distance $x$ from 
the defect, calculated numerically in the free fermion limit  
($g=1$) with $\epsilon_{0}=2$ and $n_{0}=0.25$. }
\end{figure} 
\begin{figure}[h]
\includegraphics[width=8cm]{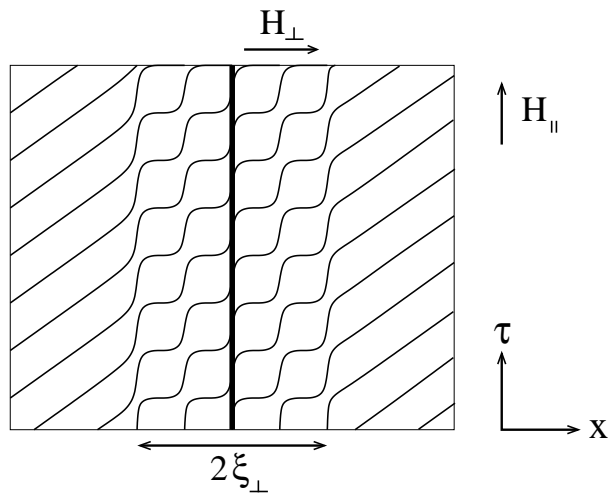}
        \caption{\label{fig:traffic_jam}
Schematic picture of the ``traffic jam'' scenario, 
for vortex lines in the vicinity of a columnar pin as described 
in the text. Because of their mutual repulsion, flux lines 
queue up in the vicinity of the pin.
}
\end{figure} 

\subsubsection{Current and Pinning Number}\label{pin_num}
Conservation of ``charge'', i.e., the number of flux 
lines in a slab, implies the existence of a 
conserved current operator, even in the presence of 
a pin and a tilt field. This is given by:
\be \hat J(x) = -i{d\hat {\cal H}(x)\over dh}=
{-i\over 2m}\left[\psi^\dagger {d\over dx}\psi 
-\left({d\over dx}\psi^\dagger \right)\psi-2h\psi^\dagger \psi \right].
\label{Jpsi}\ee
In the absence of a pin
 there is an imaginary 
current in the ground state, 
\be J=<0|J(x)|0>=
{ih\over m}<0|\psi^\dagger (x)\psi (x)|0>=ihn_0/m.\ee
Physically, this current describes vortex lines 
tilted at an angle of $\tan^{-1}(h/m)$ relative to the $\tau$-axis. 
Equivalently, each boson has an average imaginary-time ``velocity'' 
of $h/m$.
  Intuitively we might expect that 
the pin would reduce this current since vortices tend to get ``stuck'' on
it. 
However, the current is conserved and further 
describes a $\tau$-independent vortex density,  even in the 
presence of a pin.  Thus it has the same value in the vicinity 
of the pin as everywhere else in the sample. 
 Although a single pin cannot change the value 
of the current in the limit $L\to \infty$, this defect  nevertheless can have 
important finite size effects on the current which we now consider. 
We find it convenient to define a ``pinning number'', $N_p$, 
which describes the finite size reduction of the current due to the pin.
Setting $J\equiv <0|\hat J(x)|0> \equiv (ih/mL)(N-N_p)$, thus defining
\be
N_p \equiv N[J(0)-J(\epsilon_0 )]/J(0)=N+imLJ(\epsilon_0 )/h.\label{eff}\ee
Since we may think of each boson as contributing $ih/mL$ to the current in 
the clean system, $N_p$  measures the effective number of bosons which 
are not contributing to the current because they are ``stuck'' in the
vicinity of the pin. We are considering a finite size effect 
since we expect that $J(\epsilon_0 )-J(0)\propto 1/L$.  

$N_p$ may be readily calculated in the dilute limit where we can use 
the free fermion approximation. The simplest procedure is to calculate 
the ground state energy and then differentiate to get the current 
using:
\be J={-i\over L}{dE_0\over dh},\ee
which follows from Eq. (\ref{Jpsi}). 
 The ground state energy, $E_0(h)$ may be calculated by 
summing up all single particle levels below the ``Fermi surface'', as 
indicated by the prime in the summation: 
\be E_0=\sum_n{'} {K_n^2\over 2m},\ee
where the $K_n$ are given 
by Eqs. (\ref{K2}) and (\ref{phi}) for the extended states. 

We will 
assume that the limit $L\to \infty$ is being taken in Eq. (\ref{eff}) so 
that we only need the energy to O(1/L).  In this way we obtain:
\be 
J(\epsilon_0)={ihn_0\over m}+{i\over 2\pi m L}\left[
k_F\ln {k_F^2+h^2\over k_F^2+(h-m\epsilon_0)^2}
+ih\ln{(k_F+ih)[k_F-i(h-m\epsilon_0)]\over (k_F-ih)[k_F+i(h-m\epsilon_0)]}
\right].\label{Jf2}
\ee
(Here $k_F=\pi n_0$.) 
Upon extracting the pinning number, $N_p$, from Eq. (\ref{Jf2}), we have:
\be N_p = -{1\over 2\pi h}\left[
k_F\ln {k_F^2+h^2\over k_F^2+(h-m\epsilon_0)^2}
+ih\ln{(k_F+ih)(k_F-i(h-m\epsilon_0))
\over (k_F-ih)(k_F+i(h-m\epsilon_0))}\right].
\ee
Despite the fact that there is a bound state for $h<m\epsilon_0$ 
but not for $h>m\epsilon_0$, $N_p$ is a  smooth function of $h$ 
near $m\epsilon_0$  
in this large $L$ limit. (However,  a step develops 
at $h=m\epsilon_0$ in the limit $k_F\to 0$.) 
In the limit $h<<\pi n_0$, we find the asymptotic behaviors at 
large and small $\epsilon_0$:
\bea 
N_p &\to& 
{(m\epsilon_0)^2\over 2\pi^2n_0h},\ \  (m\epsilon_0<<h) \nonumber \\
&\to & {n_0\over h}\ln (|\epsilon_0 |/n_0),\ \  (m\epsilon_0>>h, n_0).
\label{pinning_number_divergence}
\eea
Remarkably, the pinning number diverges as $h\to 0$. 
This behavior can be understood in terms of the Friedel oscillations discussed 
in the previous sub-section. These oscillations imply a local density 
wave which extends out to a distance of $O(1/h)$ away 
from the pin as illustrated in 
fig.~\ref{fig:traffic_jam}.
 We may think of the particles as entering a sort of 
``traffic jam'' near the pin similar to the one which may occur 
near a toll booth. Heuristically, we think of each particle 
as waiting at the locations of the peaks in the density, which 
have spacing $1/n_0$, until the particle in front has moved ahead 
one space. The pin at $x=0$ corresponds to the toll booth.  
Unlike most real traffic jams at toll  booths, 
this one is symmetric under $x\to -x$. 
The number of particles participating in the traffic 
jam is $O(1/h)$ and represents the number of particles which are 
not participating in the current. 

It is important to note here that we have taken the limit $L\to \infty$ 
first before considering small $h$.  That is, we are assuming 
$L>>1/h$. The behavior of the current as $h\to 0$ with fixed $L$ 
is quite different, becoming linear in $h$. In this linear 
response regime the non-Hermiticity of the Hamiltonian becomes 
unimportant since the linear response to the imaginary vector 
potential can be expressed in terms of the susceptibility calculated 
at $h=0$. Apart from a factor of $i$, this is the same as the 
linear response to a {\it real} vector potential. A constant real 
vector potential of strength $-ih$ corresponds to a dimensionless magnetic 
flux, $ihL$ threading the ring. The resulting real current is 
known as the persistent current. It has been discussed for 
the case of a localized impurity potential in a Luttinger liquid, 
by Gogolin and Prokof'ev.\cite{Gogolin}   In the case of non-interacting 
fermions, these authors find an expression for the persistent current 
in terms of the transmission coefficient at the Fermi surface, $T_F$.
For the particular case of the $\delta$-function potential of 
Eq. (\ref{Hamcon}), the transmission coefficient is given by:
\be T_F={k_F^2\over k_F^2+(m\epsilon_0 )^2}.\ee
The Gogolin-Prokof'ev
 formula for the persistent current, in the limit where the 
flux goes to zero, for the case of $N$ odd [Eq. (1) of 
Ref. (\onlinecite{Gogolin})] then gives:
\be J\to {ih n_0\over m}{ \pi n_0\over m\epsilon_0}
\tan^{-1}\left({m\epsilon_0\over \pi n_0}\right) .\ee
Note that this result, linear in $h$, is independent of $L$. Naturally,  
at $\epsilon_0\to 0$, it reduces to our previous result for 
the system with no impurity. Of course, once we go beyond 
linear order the dependence of the  current on a (real) flux 
is very different than its dependence on an imaginary vector 
potential $h$. In particular the flux dependence is periodic with 
period $2\pi$ and is $O(1/L)$. 
We expect the imaginary current to cross over from the large $L$ result 
to the linear response regime when $1/h$ is of order $L$.  At this 
point the ``traffic jam'' is filling the entire system.  

In the limit $\epsilon_0\to \infty$, tunnelling of particles 
past the pin becomes very ineffective. It is then instructive to 
rederive our results using a weak tunnelling model.  This 
approach will be very useful in Sec. IV when we consider the case $g<1$. 
 In the dilute 
limit we can again analyse a non-interacting fermion problem. 
It is convenient to consider a non-interacting fermionic tight-binding model
with a weak link between sites $L$ and $1$:
\be
H=-t\sum_{i=1}^L[e^{-h}c^\dagger_ic_{i+1}+e^{h}c^\dagger_{i+1}c_i]
-\Gamma [e^{-h}c^\dagger_Lc_1+e^{h}c^\dagger_1c_L],
\label{H1}\ee
where $\Gamma <<t$ represents the weak link caused by an impurity 
with very large $\epsilon_0$. Here $c_i$ is a fermion annihilation operator.
For some purposes it is more convenient to make 
a similarity transformation to a different Hamiltonian which 
has the same (right and left) eigenvalues, chosen 
so that all the non-Hermiticity 
resides on the weak link.  This transformation is equivalent to 
the replacement:
\bea c_j&\to& c_je^{hj}\nonumber \\
c_j^\dagger &\to& c_j^\dagger e^{-hj}.\label{NUT}\eea
  This non-unitary, commutation-relation preserving 
transformation changes the Hamiltonian to:
\be
\tilde H=-t\sum_{i=1}^L[c^\dagger_ic_{i+1}+h.c.]-
\Gamma [e^{-hL}c^\dagger_Lc_1+e^{hL}c^\dagger_1c_L].
\label{H2}\ee

The lattice Schroedinger equation associated with $\tilde H$ is:
\bea 
-t(\psi_{j-1}+\psi_{j+1})&=&E\psi_j \ \  (j\neq 1,L)\nonumber \\
-t\psi_2-\Gamma e^{-hL}\psi_L&=&E\psi_1\nonumber \\
-t\psi_{L-1}-\Gamma e^{hL}\psi_1&=&E\psi_L.\label{SET}\eea
 To find the scattering states we use the ansatz:
\be \psi_j=Ae^{iKj}+Be^{-iKj},\end{equation}
where $A$ and $B$ are amplitudes and $K$ will, in general, be complex. 
Without loss of generality, we may assume ${\it Im} K\geq 0$.
The eigenvalues are then:
\be E(K)=-2t\cos K.\ee
The last two equations of  (\ref{SET}) can be rewritten as:
\bea
\Gamma e^{-hL}\left( Ae^{iKL}+Be^{-iKL}\right)&=&t(A+B)\nonumber \\
\Gamma e^{hL}\left( Ae^{iK}+Be^{-iK}\right)&=&
t\left[ Ae^{iK(L+1)}+Be^{-iK(L+1)}\right]
\label{Sch}\eea
Upon solving for $A/B$ and simplifying 
we find an  equation which determines 
the eigenvalues, i.e. the $K$'s:
\be -2\Gamma t\cosh (Lh)\sin K-\Gamma^2\sin K(L-1)+t^2\sin K(L+1)=0.
\label{K3}\ee
For  large $Lh$, we see that $K$ must have the form:
\be K_n=ih+2\pi n/L+i\phi (K_n)/L\label{K4},\ee
as before.  [We are interested in the small $\Gamma$ case, 
where a bound state occurs, so  $K_n$ is given 
by the first line in Eq. (\ref{K2}).]
 After dropping 
terms suppressed by $e^{-hL}$ (we assume $h>0$), we find:
\be -\Gamma t\sin (ih+k)+{\Gamma^2 \over 2i}e^{-h+ik+\phi}
-{t^2\over 2i}e^{h-ik+\phi}=0.\label{phi2}\ee
Here we have set $k=2\pi n/N$ and dropped the subscript $n$.  
We can now determine $\phi$ as a function of $k$. 

Once we have $\phi (k)$, we can calculate the ground state energy 
and hence the current using:
\be E_0=-2t\sum_{K} \cos K.\ee
Here we again sum over all energies whose real parts lie below the Fermi 
surface. 

Let us now just focus on the small $\Gamma$ limit. In this 
limit we may drop the second term from Eq. (\ref{phi2}).  It 
is interesting to note that this approximation corresponds to dropping the 
second term from Eq. (\ref{K3}), which in turn, corresponds 
to setting $\Gamma e^{-hL}$ to zero in the first of Eq. (\ref{Sch}). 
Thus we consider a ``one way'' model 
which ignores the hopping from $1$ to $L$ but allows it from $L$ to $1$.
Eq. (\ref{phi2}) then reduces to:
\be \Gamma e^{Lh}\sin K=\sin K(L+1).\ee
 Note that, in this small $\Gamma$ limit, $\Gamma$ and $h$ 
only appear in the combination $\Gamma e^{Lh}$.  It is therefore
convenient to define a shifted $h$ variable:
\be
e^{Lh'}\equiv \Gamma e^{Lh},\label{h'}\ee
and define $\phi '$ by Eq. (\ref{K4}) with $h$ replaced by $h'$. 
Thus:
\be \phi ' =\phi -\ln \Gamma .\ee
The dominant correction to the current, for small $\Gamma$, 
is given by using the formula for the current with no pin, 
and replacing $h$ by $h'$. 
The redefined phase shift is now determined by:
\be e^{\phi '}=1-e^{2iK}.\ee

At this point it is convenient to take the continuum limit, 
assuming that $|h+ik|<<1$, so that the phase shifts become:
\bea \phi ' &=& \ln (-2iK),\nonumber \\
\phi &=& \ln (-2iK\Gamma ).\label{phiG}
\eea
Upon comparing to Eq. (\ref{phi}), we see that Eq. (\ref{phiG}) is the 
same formula for the phase shift obtained from the original 
Hamiltonian in the limit of large $\epsilon_0$, 
with the replacement: $1/m\epsilon_0\to 2\Gamma$.
The corresponding current is:
\be J={ih'n_0\over m}+{in_0\over mL}\ln (n_0m).\ee
Upon  using Eq. (\ref{h'}), our final result for the current becomes:
\be J={ihn_0\over m}+{in_0\over mL}\ln (\Gamma n_0m).\label{JGf}\ee
Because this analysis applies in the limit $\Gamma \to 0$ we see that 
Eq. (\ref{JGf}) represents a reduction of the current. The leading dependence 
on $\Gamma$ follows immediately from the formula for 
the current with no pin upon the replacement of $h$ by 
$h'$ defined by Eq. (\ref{h'}):
\be h\to h'=h+{\ln \Gamma \over L}.\ee
The corresponding pinning number is:
\be N_p=N+imLJ(\Gamma )/h=-{n_0\over h}\ln \Gamma .\ee
This diverges logarithmically, as $\Gamma \to 0$, i.e. as 
the pinning strength goes to infinity. It agrees with the large $\epsilon_0$
result, Eq. (\ref{pinning_number_divergence}), with the identification
$\Gamma = n_0/\epsilon_0$. 

We have also calculated the pinning number via 
exact diagonalization of the noninteracting fermion tight--binding model 
(\ref{free_fermions}). [See Fig. (\ref{fig:PNf}).]
As expected from (\ref{pinning_number_divergence}), 
$N_p$ in fact grows, with decreasing $h$, 
 as $1/h$ until at $h \propto 1/L$ this divergence 
is cut off due to finite-size effects. As a result, in the linear response 
limit $L h \to 0$ the pinning number saturates at a value $N_p \propto N$ 
where $N$ is the total number of bosons. 
Another prominent feature is a steplike decrease of $N_p$ close to 
 $h_c = m \epsilon_0 $, which is a vestige of the single 
vortex depinning transition.\cite{hatano97} After being smeared 
out by interactions this step 
is only visible at low filling. 
\begin{figure}[h]
 \includegraphics[width=8cm]{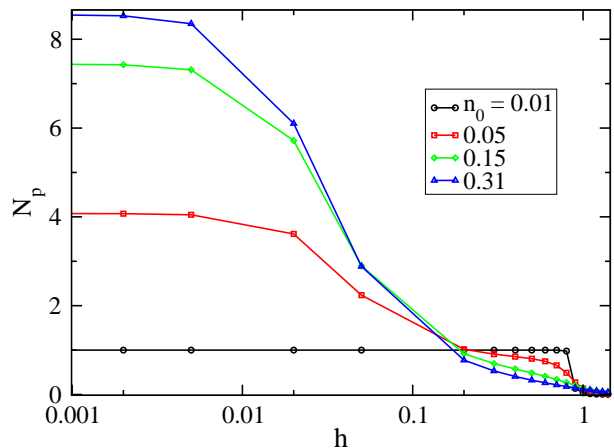}
         \caption{\label{fig:prl4}
Pinning number in the free fermion limit ($g=1$) for $L=100$ and 
$\epsilon_{0}=2$. Note the strong enhancement at small tilt $h$
and the ``step'' at $h_c \approx m \epsilon_0$ due to single--vortex depinning.
}
\label{fig:PNf} \end{figure} 
\section{Renormalization group approach to general $\lowercase{g}$}
As noted in Sec. IIC, when $h=0$, a pin is a relevant perturbation for 
$g<1$ and irrelevant for $g>1$. We use the hydrodynamic 
approach, involving  the dimensionless 
displacement field, $\theta (x,\tau )$, defined in Sec. IIB. Upon 
introducing the impurity via Eq. (\ref{F+n}), keeping  
only the most relevant parts of $\hat n(0)$, from Eq. (\ref{densu}), 
and including the tilt field from Eq. (\ref{Lthetah}), 
the Lagrangian density becomes:
\begin{equation}
{\cal L}_{\theta}={\pi \over2g}\left[{1\over c}\left( 
{\partial \theta \over \partial \tau}\right)^2
+c\left( {\partial \theta \over \partial x}\right)^2\right]
+h\partial_\tau \theta 
-\left[\epsilon_0{d\theta \over dx}+\tilde \epsilon_0 \cos (2\pi \theta ) 
\right]\delta (x),\label{Huep}
\end{equation}
where $\tilde \epsilon_0\propto \epsilon_0$, the pinning strength, but includes 
effects of eliminating short distance modes.  
The lowest order renormalization group scaling equations for 
these 2 boundary interactions are well known.  
We study the scaling of the dimensionless quantities 
$\epsilon_0$ and $\ep$, 
considering the effect of eliminating short distance 
degrees of freedom of $\theta$, thus reducing the short distance cut-off $\Lambda$. 
$\epsilon_0$  is unchanged by this reduction, 
indicating that it is a marginal coupling constant.  
 In fact, we may eliminate 
this term completely by the transformation:
\begin{equation}
\theta (x)\to \theta (x) +{\epsilon_0g\over 2\pi c}sgn (x)
\label{transf}\end{equation}
where $sgn (x)$ is the sign function with $sgn (0)=0$.
This transformation has no effect on the $\cos [2\pi \theta (0))]$  
or $\partial \theta/\partial \tau$ terms in Eq. (\ref{Huep}).
This transformation, Eq. (\ref{transf}), shifts the density of bosons by
$(\epsilon_0g/ \pi c)\delta (x)$.  Thus, the $\epsilon_0$ term in Eq. (\ref{Huep}) is an exactly 
marginal interaction.  On the other hand, the $\tilde \epsilon_0$ 
term has a $g$-dependent scaling equation. 

\subsection{$h$=0 case}
We can determine the scaling equation for $\ep$
 by observing that if we integrate out Fourier 
modes of $\theta$ with wave-vectors between $\Lambda_0$ and $\Lambda$ then 
$\cos u (x,\tau )$ gets replaced by:
\be \cos 2\pi \theta \to \left({\Lambda \over \Lambda_0}\right)^{g}
 \cos 2\pi \theta ,\ee
implying that the operator $\cos 2\pi \theta$ has a
renormalization group scaling dimension of $g$. 
Noting that the $\ep$ term in the action involves a $\tau$-integral 
but no $x$-integral, due to the $\delta (x)$ factor, we see that 
\be
\ep \to 
\ep 
\left({\Lambda \over \Lambda_0}\right)^{g-1}.\ee
Equivalently we may write the RG scaling equation:
\be {d\over dl}\ep =
(1-g)\ep ,
\ee
where $l\equiv \ln (\Lambda_0/\Lambda )$. The dimensionless pinning strength 
gets larger at longer length scales for $g<1$, but gets smaller 
for $g>1$. 

This model, Eq. (\ref{Huep}), has been well-studied in the closely related context 
of a quantum fermion system\cite{Kane} or 
a quantum spin chain\cite{Eggert}  with a point impurity, from 
which it arises by bosonization. In these contexts it has 
been rather well established by numerical and analytic work 
that, for $g<1$, starting even with a small impurity strength 
the long distance (low energy) behavior is that of a 
``cut chain'' with the large impurity strength effectively 
decoupling the two sides. On the other hand, for $g>1$, starting 
even with a large impurity strength the long distance behavior 
is that of a ``healed'' chain with no impurity. 
 We believe that this 
is also the case for the bosonic version of the model, 
defined by Eq. (\ref{Hamcon}). 

The cut chain fixed point is easily studied 
in the phase boson representation. A boundary 
condition, $\theta (0)=0$ must be imposed.  Note 
that we should really think of the $x>0$ and $x<0$ 
parts of the systems as being independent in this 
limit.  (We take $L\to \infty$ for this discussion.)
Thus we get two boundary conditions, 
$\theta (0^{\pm})=0$. These imply $\partial \theta /\partial t
\propto \Pi_\theta =0$ 
and hence, from Eq. (\ref{Piu}), the Neumann boundary condition 
on $\phi$: 
\begin{equation}
 {d\phi \over dx}(0^{\pm}) =0.\label{Neum}\end{equation}
This boundary condition modifies the correlation functions. 
Of course any correlation function of two fields on opposite sides 
of the pin is zero.  The correlation function of two fields 
on the same side is also modified.
One way of calculating these correlation functions, with 
the boundary condition, is to decompose the free boson fields 
$\phi$ into left and right moving components:
\begin{equation}
\phi (x,t)=\phi_L(t+x/c)+\phi_R(t-x/c)\end{equation}
(The most general solution of the equations of motion, 
$(\partial_t^2-c^2\partial_x^2)\phi=0$ can be decomposed in this way.)
Eqs. (\ref{Piu}) and (\ref{canu}) then imply: 
\begin{equation}
\theta  (x,t)={g\over \pi}[\phi_L(t+x/c)-\phi_R(t-x/c)].\end{equation}
The boundary condition can thus be written:
\begin{equation}
\phi_L(t,0^{\pm})=\phi_R(t,0^{\pm}).\label{bcLR}\end{equation}
Let us focus on correlations for $x>0$, for example. 
Then the boundary condition of Eq. (\ref{bcLR}) implies, 
since $\phi_{R,L}$ is a function of $t\mp x/c$ only, that 
we may regard $\phi_R(x)$ as the analytic continuation of 
$\phi_L(x)$ to the negative axis:
\begin{equation}
\phi_R(x)\equiv\phi_L(-x),\ \  (\hbox{for}\ x>0).\end{equation}
This has the effect of making the density and boson creation operators bi-local:
\begin{eqnarray}
\psi^\dagger (x)&\approx &\sqrt{n_0}e^{i[\phi_L(x)+\phi_L(-x)]}\nonumber \\
n(x)-n_0&\approx &{2g\over \pi} {d\phi_L\over dx}+
 \hbox{constant}\times \cos \{ 2\pi n_0 x+
2g[\phi_L(x)
-\phi_L(-x)]\} .\label{psinbc}\end{eqnarray} 
The correlation functions can now be calculated using:
\begin{equation}
<\phi_L(x,\tau )\phi_L(y,0)>=-{1\over 4g}\ln [
(x-y)+ic\tau]+\hbox{constant}
\end{equation}
Thus the correlation function of the boson creation operator, 
discussed in Sec. IIC, becomes:
\begin{equation}
<\psi^\dagger (x,\tau )\psi (y,0)>\propto 
\left\{xy\over [(x-y)^2+c^2\tau^2][(x+y)^2+c^2\tau^2]\right\}^{1/4g},
\label{psicorimp}\end{equation}
where $x$ and $y$ are on the same side of the pin. 
Note that in the limit $x,y>>|x-y|,c|\tau |$ we recover the bulk 
behavior [Eq. (\ref{psicorr})]:
\begin{equation}
<\psi^\dagger (x,\tau )\psi (y,0)>\propto 
\left\{ 1\over  [(x-y)^2+c^2\tau^2]\right\}^{1/4g}.\end{equation}
On the other hand, in the limit $c|\tau |>>x,y$ we obtain 
the ``boundary critical behavior'':
\begin{equation}
<\psi^\dagger (x,\tau )\psi (y,0)>\propto 
\left\{ 1\over  |\tau |\right\}^{1/g}.\end{equation}
Thus we see that the operator, $\psi^\dagger$, which has a bulk 
scaling dimension of $1/4g$ has a boundary scaling dimension 
which is twice as big, $1/2g$. 
To understand this result, note 
 that, without the boundary condition, $\phi_L$ 
and $\phi_R$ are independent fields, so that 
both factors $e^{i\phi_L}$ and $e^{i\phi_R}$ contribute 
equal amounts $1/8g$ to the scaling dimension of $\psi^\dagger$. 
After imposing the boundary condition $\psi^\dagger (0)$ 
becomes the operator  $e^{2i\phi_L(0)}$ which has dimension $1/2g$. 

When the pin is relevant, $g<1$, we may calculate the density 
oscillations at long distances from the pin 
by assuming that $\ep \to \infty$ and using the Neumann boundary 
condition of Eq. (\ref{Neum}). This constraint leads to Eq. (\ref{psinbc}) which 
leads to:
\be <0|n(x)|0>\to n_0 + {\hbox{constant}\cdot \cos (2\pi n_0x)
\over |x|^{g}}.\ee
For an irrelevant pin, we expect the result of 
lowest order perturbation theory in $\epsilon$ to 
be valid at long distances, giving Eq. (\ref{dnPS}). Note
this involves a different (larger) exponent than the one 
which occurs for a relevant pin. 
These density oscillations are the bosonic version of 
the generalized Freidel oscillations discussed 
for fermionic systems and spin chains in Ref. 
(\onlinecite{Friedel,Meden02}). 

We can now study the stability of the cut chain 
fixed point. As discussed above, in this limit the system 
decouples into two separate sections to the left and right of 
the pin.
 If $\ep$ is very large but finite, there will be 
a (dimensionless) 
weak tunnelling matrix element, $\Gamma \propto 1/\ep$, between 
the two sides. In the $\psi$ representation, this effective 
Hamiltonian is:
\begin{equation}
\hat H=\hat H_{-}+\hat H_{+}-\Gamma c[\psi^\dagger (0^-)\psi (0^+)+h.c.]
\label{Heff}\end{equation}
Here 
\begin{equation}
\hat H_+=\int_{x>0} dx{1\over 2m}{d\psi^\dagger \over dx}
{d \psi \over dx}+\hbox{constant}*\psi^\dagger (0)\psi (0)
+{1\over 2}\int_{x,y>0} dx dy \hat n(x)V(|x-y|)\hat n(y),
\end{equation}
 $H_-$ is defined 
similarly.  (Alternatively, an interaction term similar to that in 
Eq. (\ref{Hamlat}) may 
be maintained between the two sides.  The important thing 
is that there is no motion of bosons across the pin in this limit.)
This is conveniently written in terms of the phase boson, $\phi$ 
 with 
the boundary condition $d\phi /dx =0$ at $x=0^\pm$.  Thus we get two 
copies of the Lagrangian of Eq. (\ref{L-phi})  for 
$x> 0$ and $x<0$. As explained above, the tunnelling term, analogous 
to a Josephson coupling across the weak link,  becomes:
\be -\Gamma c\psi^\dagger (0^-)\psi (0^+)+h.c.
\propto \Gamma \cos 2[\phi_L(0^+)-\phi_L(0^-)],\ee
of scaling dimension $1/g$. (Because $\psi (0^+)$ and $\psi (0^-)$ 
become independent operators upon imposing the boundary condition, 
their scaling dimensions simply add. )  Thus, the renormalization 
group equation obeyed by $\Gamma$ is:
\be {d\over dl}\Gamma 
=\left( 1-{1\over g}\right)\Gamma + \ldots 
\label{GammaRG}\ee
We conclude that a weak tunnelling across the pin is relevant 
for $g>1$ but irrelevant for $g<1$, i.e.  precisely 
the inverse of the situation for a weak pinning potential, $\ep$. 
This implies consistency of the bold assumptions that a 
weak $\ep$ will renormalize all the way to the broken 
chain fixed point (corresponding to $\Gamma 
\to 0$) for $g<1$ and that even a large $\ep$ will renormalize to 
$0$ for $g>1$.
We present numerical evidence to verify this conjecture 
in Appendix C and D. 

\subsection{Density oscillations for $h>0$}
We now consider the effect of a non-zero transverse field, $h$, 
in Eq. (\ref{Huep}).  We may eliminate the term in Eq. (\ref{Huep})
proportional to $h$ by a shift of the $\theta$ field:
\be \theta (x,\tau )\to \theta (x,\tau )-(gch/\pi )\tau .\label{shift}\ee
Let us first consider the density 
oscillations in the limit of a weak pin, in lowest 
order perturbation theory in $\ep$ as in Sec. IIC.  The 
singular part of the density-density correlation function
picks up an extra $h$-dependent phase from this shift:
\begin{equation}
<n(x,\tau )n(0)>\to n_0^2+{gc^2\over 2\pi^2}{c^2\tau^2-x^2\over 
(c^2\tau^2+x^2)^2}
+\sum_{p=1}^\infty 
 {A_p\cos [(2\pi n_0x-2ghc\tau )p]\over [x^2+c^2\tau^2]^{gp^2}}+\ldots ,
\end{equation}
Upon Fourier transforming, we see that the singularities of the 
structure function have moved off the $q_x$ axis to 
$(q_x,q_\tau )=(2\pi n_0,2ghc)p$, as shown in Fig. (\ref{fig:Shn0}). 
The linear response to a static pin is proportional to 
$S(q_x,0)$, [see Eq. (\ref{delta_n})], which is non-singular, 
as indicated in Fig. (\ref{fig:Shn0}).  Focussing on what 
was the leading singularity, at $q_x=2\pi n_0$, we find:
\be <0|\hat n(x)|0> \approx n_0+
\hbox{constant}\cdot \ep\cdot \cos (2\pi n_0x)
\int d\tau {e^{i2ghc\tau}\over [x^2+c^2\tau^2]^{g}}.
\label{densirrh}\ee
This integral may be expressed in terms of the modified 
Bessel function of the second kind, $K_{g-1/2}$:
\be 
\int d\tau {e^{i(2\pi hn_0/m)\tau}\over [x^2+c^2\tau^2]^{g}}
={2\over c\xi^{2g-1}\Gamma (g)}\left( {\xi\over 2|x|}\right)^{g-1/2}
K_{g-1/2}(x/\xi ),\ee
with characteristic length scale
\be \xi = cm/(2\pi n_0h)=1/(2gh)\ee
and where $\Gamma (g)$ is Euler's $\Gamma$ function. 
In the limit $|x|<<\xi$, we obtain the zero tilt result, 
Eq. (\ref{dnPS}). In the opposite limit, $|x|>>\xi$, 
we obtain exponentially screened Friedel oscillations:
\be <0|\hat n(x)|0> \approx n_0+
\hbox{constant}\cdot \ep\cdot \cos (2\pi n_0x)e^{-|x|/\xi}{1\over 
\xi^{g-1}|x|^{g}}.
\label{densirrh2}\ee

Beyond the new characteristic  
length scale, $\xi = 1/(2gh)$, we expect that the pin loses its effectiveness in ordering the 
vortices, for {\it any} value of $g$. It is natural to assume that this new length scale 
acts as an infrared cut-off on the renormalization of 
the pinning strength, $\tilde \epsilon (l)$. Thus 
the long distance physics should be controlled by: 
\be
\tilde \epsilon_0 (h)\approx \tilde \epsilon_{00} 
(\Lambda_0/h)^{1-g}.\ee
Here $\tilde \epsilon_{00}$ is the ``bare'' value of 
$\ep$, that is, before renormalization. Provided that 
$\tilde \epsilon_{0}(h)<<1$, we expect the perturbative 
result to be valid at long distances. 
Thus it would  be valid for $g>1$ even 
if the bare dimensionless pinning potential is not small, for 
sufficiently small $h$. It would also be valid for $g<1$ provided 
that the bare dimensionless pinning potential is sufficiently 
small and $h$ is not too small. As we showed in the previous 
section, for the case $g=1$, corresponding to free fermions, 
this exponential decay of the Friedel oscillations at 
long distances holds for any $\ep$. Even if the above 
criteria are not satisfied we still expect exponentially 
decaying density oscillations at long distances for any 
values of $g$ and $\ep$ provided that $h\neq 0$. 
\begin{figure}
\begin{center}
\includegraphics[width=0.4\linewidth]{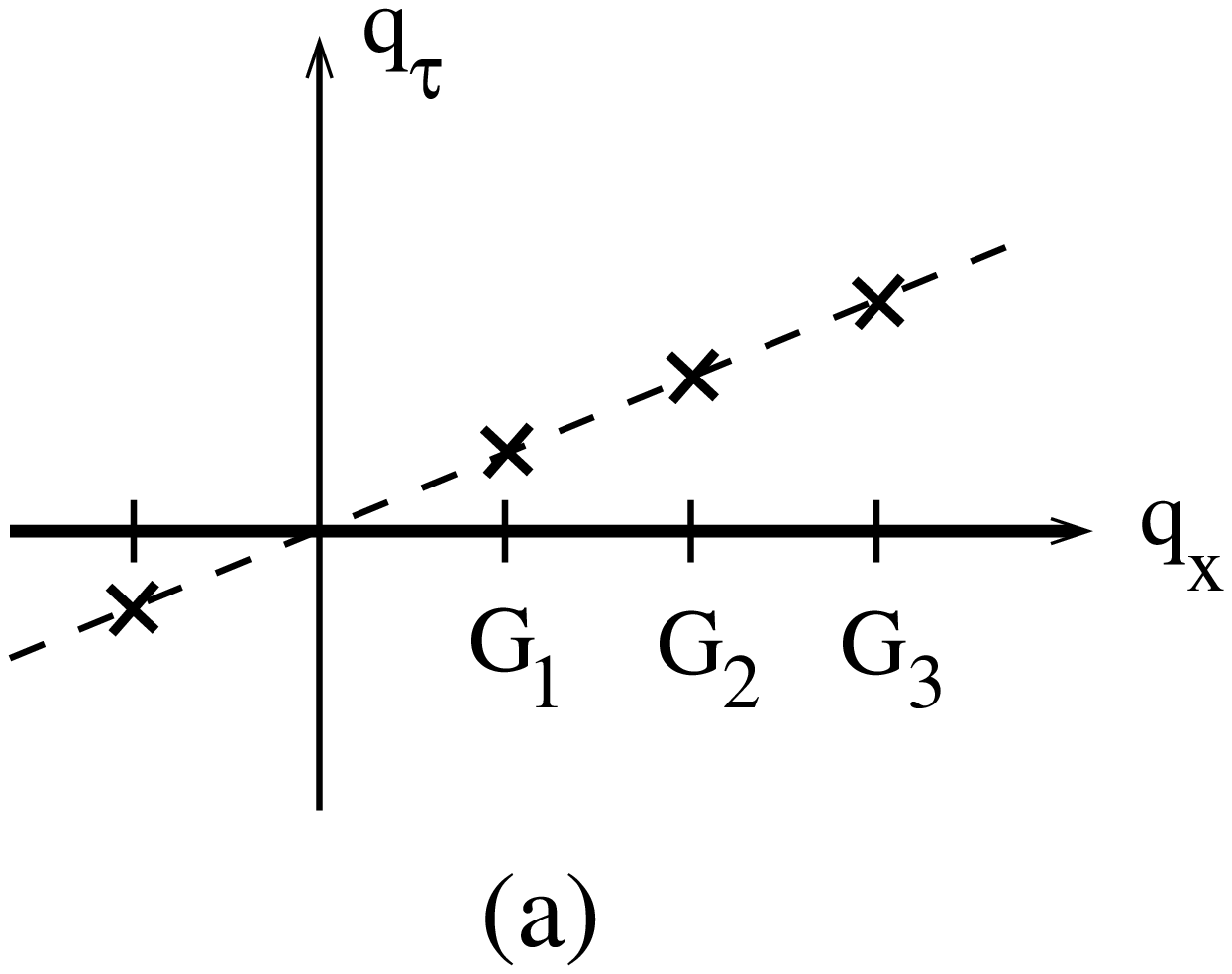}
\hspace{0.1\linewidth}
\includegraphics[width=0.4\linewidth]{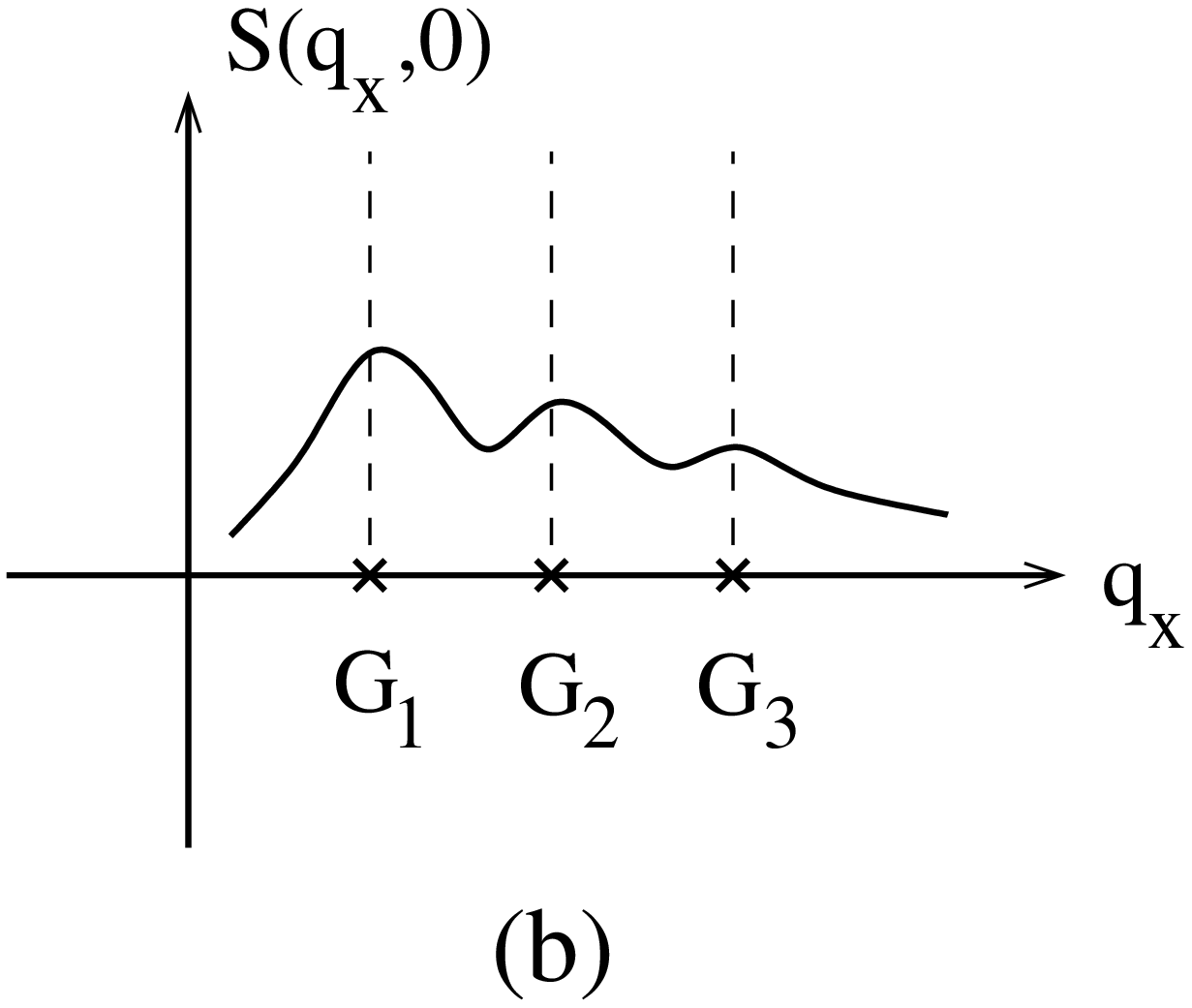}
\caption{(a) Reciprocal lattice vectors in the $(q_x, q_\tau)$--plane for 
$h \ne 0$.  In contrast to Fig. (\ref{fig:S}),
 the singularities now occur off the
$q_x$--axis, where the potential due to the columnar pin is zero.  
(b)  As shown in this profile, the structure function $S(q_x, 0)$ is now 
finite and nonsingular everywhere along the $q_x$--axis.}
\label{fig:Shn0}
\end{center}
\end{figure}

DMRG results for the Friedel oscillations with $g<1$ are shown in 
fig.~\ref{fig:density_general}. 
Note that, as in the free Fermion limit $g=1$, a finite tilt $h$ 
tends to strongly suppress the oscillations. 
Our data are consistent with a crossover from power--law to exponential decay 
with increasing $h$, although the system sizes are too small 
to extract precise exponents.  

\begin{figure}[h]
 \includegraphics[width=8cm]{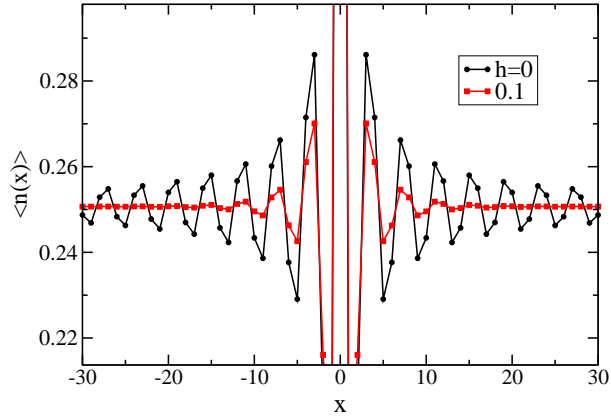}
         \caption{\label{fig:density_general}
Friedel oscillations calculated by DMRG for $\epsilon_0=2$, 
$U=10$, $V=4$, $L=128$ and $n_0=0.25$, corresponding to $g\approx 0.72$.
}
 \end{figure}

\subsection{Current}
We first consider the case $h\to 0$ for fixed $L$.
As discussed in Sec. III, when $hL<<1$, the current becomes linear 
in $h$ and proportional  to the (persistent) current that results 
from a real vector potential. This real persistent current 
was analysed, for arbitrary $g$, 
by Gogolin and Prokof'ev \cite{Gogolin} and we may simply take over their results, 
replacing the dimensionless magnetic flux, $\varphi$, by $ihL$. In 
the case of a relevant pin,  $g<1$, Eq. (18) of [\onlinecite{Gogolin}] gives:
\be J\propto {ih\over L^{1/g-1}}.\label{JLR}\ee
We see that the linear response current vanishes as a non-trivial 
power of $1/L$, whereas it is independent of $L$ for $g=1$.
As observed in [\onlinecite{Gogolin}] 
the $L$-dependence of the current
 can be understood from the renormalization group behavior 
of the effective weak tunneling matrix element, $\Gamma$, 
 given by Eq. (\ref{GammaRG}),
\be \Gamma (L)\propto L^{1-1/g}.\ee
In the weak tunnelling limit of the non-interacting case ($g=1$),
 the transmission coeffient 
$T_F\propto |\Gamma |^2$ 
and the persistent current $J \propto \sqrt{T_F}\propto \Gamma$.  For 
$g<1$, replacing $\Gamma$ by $\Gamma (L)$, leads to  (\ref{JLR}).  In the other case of an irrelevant pin, 
$g>1$, the transmission coefficient renormalizes to $1$ at low 
energies and long distances so we expect to recover the result 
for the system with no pin, namely
\be J\to {ih\over m}n_0.\ee
(Note that we require not only $1/h>>L$ but also that $L$ 
is much greater than the characteristic length scale required to send 
 $ \ep \to 0$. )

In fig.~\ref{fig:linear_current} we show DMRG results for the 
imaginary current in the case of a relevant pin 
for system sizes up to $L=256$.  
The finite--size scaling (\ref{JLR}) of the linear--response current 
is confirmed with high accuracy in the limit $hL\leq 1$.  
\begin{figure}[h]
\includegraphics[width=8cm]{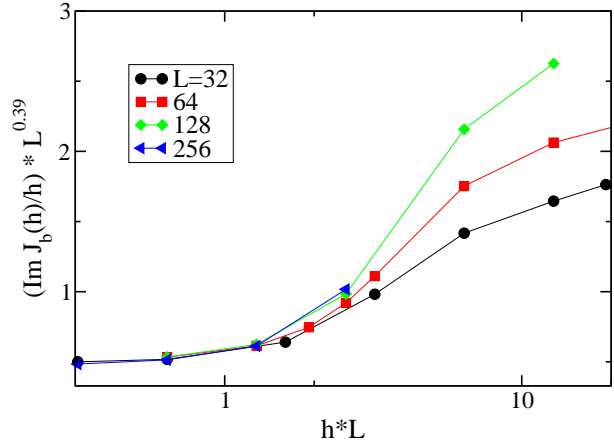}
        \caption{\label{fig:linear_current}
Finite--size scaling of the current (DMRG results) for filling 
$n_0 = 0.25$, $\epsilon_0 = 2$ and a relevant pin ($g\approx 0.72$). 
Note the data collapse in the linear--response regime $hL \to 0$.  
}
\end{figure}

We now consider the other limit, $L>>\xi \equiv 1/2gh$, where 
we can characterize the current in terms of the pinning number, $N_p$. 
In the case, $g>1$, where the pin is irrelevant, it is reasonable 
to calculate the correction to the current due to the pin 
(which can be expressed in terms of the pinning number) 
in lowest order perturbation theory in $\ep$, using 
the bosonized $\theta$-representation. We again start with
the Lagrangian density of Eq. (\ref{Huep}).

In second order perturbation theory in $\ep$, 
the leading correction to the ground state energy is proportional to:
\begin{equation}
\delta E_0 \propto 
\ep^2\int_{-\infty}^\infty d\tau  <
\cos [2\pi \theta (0,\tau )]\cos [2\pi \theta (0,0)]>.\label{corr}
\end{equation}
Note that this ground state energy is given by the logarithm of the 
classical partition function associated with Eq. (\ref{Huep}). 
Upon taking into account the shift of Eq. (\ref{shift}), we find:
\be \delta E_0\propto \ep^2\int_{\tau_0}^{\infty}{d\tau \over |\tau|^{2g}}
\cos [2gch\tau ],\label{intJ}\ee
where $\tau_0$ is a short distance cut-off. 
 For $2<2g<3$, 
Eq. (\ref{intJ}) leads to:
\be \delta E_0 \propto \ep^2 \left[ -h^{2g-1}+\hbox{constant} \cdot 
\tau_0^{-2g+1}\right] .\ee
The correction to the current due to the pin, is thus:
\bea \delta J &=& {-i\over L}{d\delta E_0\over dh}\propto {i\over L}
\ep^2h^{2g-2}\nonumber \\ 
&\propto& {i\over L}\ep (h)^2.\eea
The last entry expresses $\delta J$ in terms of the renormalized 
pinning strength at the length scale $1/h$. 
After setting $g=1$, we find agreement with the exact result in the dilute 
limit, if we also assume that $h<<\pi n_0$. We  
only expect the phonon representation to be valid when 
$h<<\pi n_0$ since $n_0$ represents an effective ultra-violet cut off  
and we have thrown away higher derivative terms in deriving 
this effective Lagrangian. 
For $2g>3$, the leading behavior at small $h$, from Eq. (\ref{intJ}), 
is:
\be \delta J \propto {i\over L}\ep^2 h/\tau_0^{2g-3}.\ee
Thus the pinning number behaves as:
\bea N_p &\propto& {\ep^2\over h^{3-2g}}\ \  (2<2g<3)\nonumber \\
&\propto& \ep^2 \ \  (2g>3),\label{N_P}\eea
results applicable in the limit $Lh\to \infty$. 

We may straightforwardly extend this calculation to finite $Lh$.  
  Then we must use the finite $L$ 
version of the correlation function in Eq. (\ref{corr}), 
which can be obtained by a conformal transformation 
and effectively
replaces $\tau$ by $(L/\pi c)\sinh (\pi c\tau /L)$ in 
Eq. (\ref{intJ}). Upon rescaling the $\tau$ integration variable, $u\equiv \pi c\tau /L$, 
and using Eq. (\ref{cg}), we obtain:
\be \delta E_0\propto \ep^2(L/\pi c)\int_{1/(\pi c\tau_0/L)}^{\infty}{du 
\over |(L/\pi )\sinh u|^{2g}}
\cos [2ghLu/\pi],\label{intJ2}\ee
Differentiating with respect to $h$ and dividing by $h$ gives 
the pinning number:
\be
N_P\propto \ep^2 L^{2-2g}h^{-1}\int_0^\infty du{u \sin (2ghLu/\pi )\over 
\sinh^{2g} u} ,\ \  (1<g<3/2).\ee
Note that we have set the lower limit of the integral to $0$ since 
it converges.  Thus we obtain the scaling form:
\be N_p=L^{3-2g}f(Lh),\ee
where:
\be f(x)\propto {1\over x}\int_0^\infty du {u \sin (2gux/\pi )
\over (\sinh u)^{2g}}.\ee
We see that $f(x)\to x^{2g-3}$ as $x\to 0$. 
Using DMRG, we have numerically checked this scaling 
for the case $2g<3$ and have found 
good agreement (see fig.~\ref{fig:scaling_function}).

\begin{figure}[h]
\includegraphics[width=8cm]{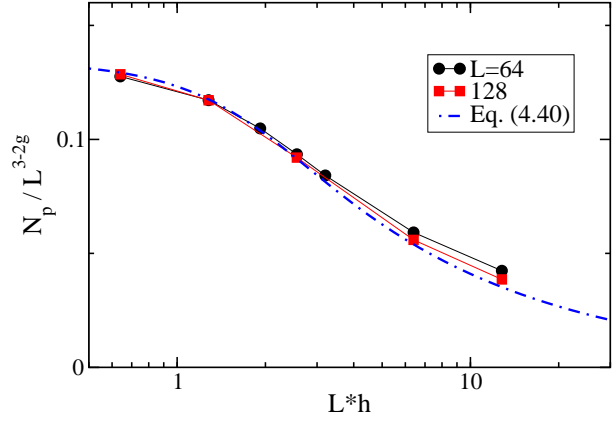}
        \caption{\label{fig:scaling_function}
Scaling function for the pinning number in the case of an irrelevant pin 
($U=10$, $V=0$, $g\approx 1.19$) with boson density $n_0=0.25$ and 
pinning potential $\epsilon_0 = 2$.
}
\end{figure} 
To study the case of a relevant pin, $g<1$, 
we use the weak tunnelling model of Eq. (\ref{Heff}).  
We  determined the effective $\Gamma$ parameter by using the 
$h=0$ RG equations of Eq. (\ref{GammaRG})  out to a length scale of $1/h$:
\be \Gamma \propto h^{1/g -1}.\label{Gamma}\ee
We noted in Sec. III that 
it is convenient to make a similarity transformation of
the Hamiltonian such that the hopping terms have the form of Eq. (\ref{H2})
in which all the 
non-Hermiticity resides on the weak link. We then 
showed that the exponentially small hopping term, $\propto \Gamma e^{-hL}$ 
 from 
 right to left can be dropped at small $\Gamma$, keeping 
only the exponentially large hopping term, $\propto \Gamma e^{hL}$ from 
left to right.  Thus we have a ``one-way hopping model''.  
It seems plausible that this approximation can also be 
made away from the $g=1$ free fermion case, as long as 
the effective $\Gamma$ is very small.  Once we make this approximation, 
$\Gamma$ and $h$ appear {\it only} in the combination 
$\Gamma e^{hL}$. Then, since we expect that, in the infinite $L$ 
limit the current goes as: $i{n_0 \over m}h$,  the finite 
size correction at small $\Gamma$ must go as:
\be J\approx i{n_0 \over m}\left[h+{\ln \Gamma \over L}\right].
\ee
We now replace $\Gamma$ by its value in Eq. (\ref{Gamma}) 
which leads to:
\be J\approx i{n_0 \over m}\left[h+{1 \over L}
\left\{ \left({1\over g}-1\right)
 \ln h + \hbox{constant}\right\} \right].\ee
Thus the pinning number is:
\be
N_p=N_b{J(\ep =0)-J(\ep )\over J(\ep =0)}={n_0 \over h}
\left[ \left({1\over g}-1\right)
 |\ln h | + C \right],
\label{log_conjecture}
\ee
where $C$ is  a non-universal constant.  Note 
that the universal number $g$ appears here as an amplitude, not as 
an exponent. 
We emphasize that we expect this amplitude 
to be universal by this argument, although the sub-dominant 
constant is not. Thus a relevant pin only increases the 
pinning number by a factor of $|\ln h|$ compared to 
the free fermion (low vortex density) case where the pin is marginal. 

Our DMRG results for $N_p$ in the $g<1$ case, shown in 
fig.~\ref{fig:log_conjecture}, confirm these analytic predictions. 
The logarithmic behavior (\ref{log_conjecture}) is clearly observed 
for two different values of the Luttinger liquid parameter $g$. 
Note that at $Lh \propto 1$ the divergence of $N_p$ is cut off, 
analogous to the free--fermion limit. 

\begin{figure}[h]
\includegraphics[width=8cm]{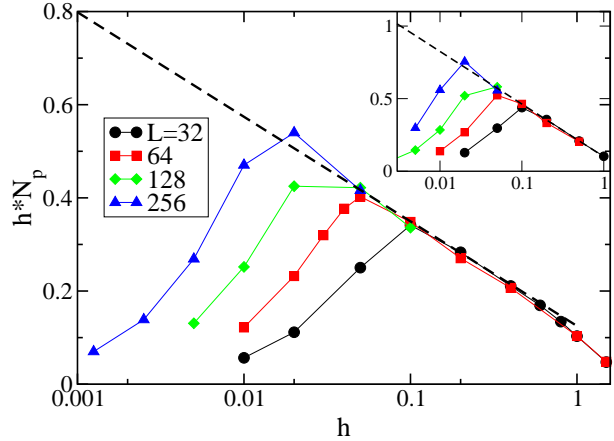}
        \caption{\label{fig:log_conjecture}
DMRG results for the pinning number.
Main plot: filling $n_0=0.25$, 
$\epsilon_0 = 2$ and a relevant pin ($U=10$, $V=4$, corresponding to 
$g\approx 0.72$). 
The dashed line gives the logarithmic behavior in Eq. (\ref{log_conjecture}) 
with $C=0.5$. 
Inset: same for $U=20$, $V=10$, $g\approx 0.62$ and offset $C=0.4$. 
}
\end{figure} 

In Appendix F we briefly discuss the phase-boson representation 
of the weak tunnelling model after the similarity transformation, 
Eq. (\ref{H2}), for general $g$, pointing out its connection 
with a model of current interest in string theory.  

\section{Point Disorder}
This paper focuses on the physics of thermal fluctuations in very
clean (1+1)--dimensional superconducting slabs.  The appropriate
physical conditions necessary to neglect weak point disorder due to
oxygen vacancies, proton irradiation, etc.\cite{blatter94} in high $T_c$ 
superconductors are discussed in Appendix E.  We can, however, get
some insight into the influence of a single extended defect in the
presence of strong point disorder using the linear response formalism
discussed earlier. A treatment of a single columnar pin in the presence 
of strong point disorder, including a full renormalization group analysis, 
will appear in a future publication.\cite{polkovnikov}

The effect of a quenched random distribution of point pins on vortex
arrays in thin superconducting slabs has been discussed in a number of
publications\cite{fisher89a,hwa93,hwa94}. In the presence of point disorder, the
background elastic free energy (\ref{elastic_energy})
 on which we impose a single
columnar pin becomes 
\bea
&& \int dx d\tau \big\{\frac{1}{2} c_{11} (\partial_x u)^2 + 
\frac{1}{2} c_{44} (\partial_\tau u)^2 - \mu_x (\partial_x u) 
- \mu_\tau (\partial_\tau u) \nonumber \\
&& - h (\partial_{\tau}u) + V_0(x,\tau)\, 
\cos\left[2\pi u(x,\tau) / a_0 - \beta(x,\tau)\right]\big\}. 
\label{elastic_w_pin}
\eea
As pointed out in Ref. \onlinecite{fisher89a}, the statistical mechanics associated with
Eq.~(\ref{elastic_w_pin}) is similar to that of an anisotropic two dimensional random
field XY model with, however, no topological defects such as the
dislocations shown in Fig.~(\ref{fig:top_defect}).  
As in charge density wave physics\cite{lee}, the cosine term acts 
to locally fix the phase of the (1+1)--dimensional vortex crystal at $\beta(x,\tau)$ 
with a strength determined by the amplitude $V_0(x,\tau)$.  
The couplings $\mu_x(x,\tau)$ and $\mu_\tau(x,\tau)$ are
zero mean random variables which describe local variations in the
preferred density and tilt of the vortex lines induced by the
particular configuration of point disorder.  At long wavelengths, 
$\mu_x(x,\tau)$ and $\mu_\tau(x,\tau)$ 
renormalize in the same way and can be taken to be Gaussian random
variables described by a single variance $\sigma$\cite{fisher89a,hwa94}, 
\be
\overline{\mu_i(x,\tau) \mu_j(x',\tau')} = \sigma\, \delta_{ij}\,  
\delta(x-x')\, \delta(\tau - \tau'),
\ee
where  $i,j=x,\tau$ and the overbar represents a quenched average 
over the disorder. We also take  $V_0(x,\tau)$ to be given by 
a Gaussian distribution, with variance
\be
\overline{V_0(x,\tau) V_0(x',\tau')} = \Delta_0\, \delta(x-x')\, 
\delta(\tau - \tau')
\label{disorder_correlator}
\ee
while  $\beta(x,\tau)$ is uniformly distributed on the interval $[0,2\pi]$.

When averaged over point pinning, the linear response 
equation (\ref{delta_n_first}) becomes 
\be
\overline{\delta n(x,\tau)} = -\int dx' d\tau' \frac{V_D(x')}{T}\, 
\overline{C(x-x', \tau-\tau')},  
\ee
with 
\be
C(x-x', \tau-\tau') = \overline{<n(x,\tau) n(x',\tau')>_0} - 
\overline{<n(x,\tau)>_0}\, \overline{<n(x',\tau')>_0}. 
\ee
The Fourier space version of these relations reads
\be
\overline{\delta n(q_x,q_\tau)} = \overline{S(q_x, q_\tau)}\, 
\frac{L_\tau \delta_{q_\tau, 0}}{T}\, \hat{V}_D(q_x),
\ee
where
\be
\overline{S(q_x,q_\tau)} = \overline{<|n(q_x, q_\tau)|^2>},
\ee
and all averages are evaluated in the {\it absence} of the columnar defect. 

We first set $h = 0$ and apply the renormalization group analysis of the
bulk statistical physics problem defined above to evaluate 
$\overline{S(q_x, q_\tau)}$ and hence determine the change 
in vortex density due to a single columnar pin.
The recursion relations for the scale-dependent 
couplings $g(l)$, $\sigma(l)$ and $\Delta_0(l)$ 
are\cite{hwa93,hwa94,cardy,toner}    
\be
\frac{dg(l)}{dl} = 0,
\ee
\be
\frac{d\sigma(l)}{dl} = C_1 \Delta_0^2(l)
\label{sigma_flow}
\ee
\be
\frac{d\Delta_0(l)}{dl} = 2 [1 - g(l)] \Delta_0(l) - 
C_2 \Delta_0^2(l)
\ee
where $C_1$ and $C_2$ are positive constants.  
The renormalization group flows in the $(g,\Delta_0)$--plane are shown 
in Fig.~\ref{fig:RGpoint}.   
\begin{figure}
\begin{center}
\includegraphics[width=0.45\linewidth]{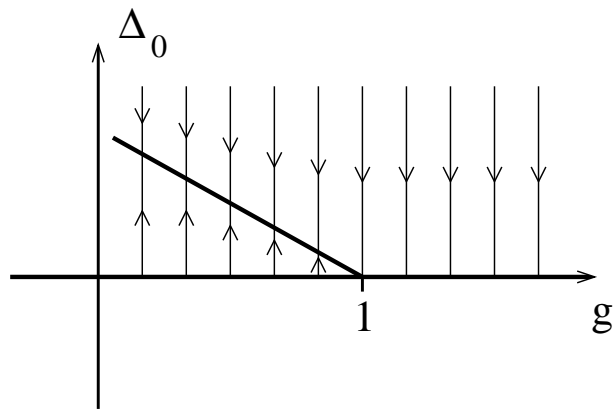}
\caption{Renormalization group flows in $g$ and $\Delta_0$ for the random 
field 
XY model which describes (1+1)--dimensional vortex arrays subjected to 
point disorder.\cite{cardy}  The Luttinger liquid fixed line at $\Delta_0 = 0$ is 
is stable to point disorder for $g > 1$, which is also the regime where a 
single columnar pin is irrelevant.  This line becomes unstable to a new 
fixed line where point disorder is important for $g < 1$.}
\label{fig:RGpoint}
\end{center}
\end{figure}

The line of fixed points at $\Delta_0=0$ in Fig.~\ref{fig:RGpoint}  
is \emph{stable} to point disorder when $g>1$.  
This is also the regime where columnar pins are irrelevant in pure
systems.  However, as the point disorder strength tends to zero, it
generates via Eq.~(\ref{sigma_flow}) a nonzero variance for the couplings 
$\mu_x(x,\tau)$ and $\mu_\tau(x,\tau)$ in
Eq.~(\ref{elastic_w_pin}). After setting the cosine coupling to zero in 
Eq.~(\ref{elastic_w_pin}) it is straightforward to show\cite{rubinstein}  
that the leading singular term ($m=1$) in
 Eq.~(\ref{etadef}) is replaced by
\be
\overline{<n(x,\tau) n_(0,0)>} -n_0^2\sim 
{\cos (2\pi n_0x)\over [x^2 + c^2 \tau^2]^{\eta/2}}, 
\ee
with 
\be
\eta = 2g + \frac{\sigma_\infty}{2\pi}, 
\ee
where $\sigma_\infty = \lim_{l\to \infty} \sigma(l)$ 
is a (positive) correction to the Luttinger liquid result.  Hence, $\eta>2$ 
for all $g \ge 1$ and we conclude that $\overline{S(q_x,q_\tau)}$ 
never diverges near any of the reciprocal lattice vectors. 
In particular, $\overline{S(q_x,q_\tau)}$ never diverges at $G_1$, 
suggesting that an isolated columnar pin becomes even 
\emph{more} irrelevant for in the presence of
strong point disorder when $g>1$.

A new stable line of fixed points appears in Fig.~\ref{fig:RGpoint} 
for $g\le 1$, signaling the onset of a ``vortex glass'' phase in 
(1+1)--dimensions\cite{fisher89a}. 
The analysis of Hwa and Fisher\cite{hwa94} (see also, Ref.~\onlinecite{toner}) 
shows that 
\be
\overline{\langle n_1(x,\tau) n_1^*(0,0)\rangle} \sim 
\exp\left[-C_3 \ln^2(x^2 + c^2 \tau^2)\right]
\ee
where $C_3$ is a positive constant. 
Because this decay is faster than any power law, $\overline{S(q_x, q_\tau)}$ 
is again finite everywhere along the $q_x$--axis, 
suggesting that an isolated columnar defect is
also asymptotically irrelevant in the presence of strong point
disorder in this regime.  However, if the point disorder is weak, the
effective pinning strength of the columnar defect can grow quite large
over intermediate length scales.  The subtle and complex physics which
distinguishes the response of (1+1)--dimensional vortex arrays to a
columnar defect above and below the vortex glass transition will be
discussed in Ref.~(\onlinecite{polkovnikov}).

We conclude this brief discussion of point disorder with two comments:
The tilt field $h$ is a strongly relevant variable in (1+1)--dimensions
(the recursion relation for $h$ reads\cite{hwa93} $dh(l)/dl =  h(l)$) 
for both pure and disordered systems.  
When point disorder is present, the structure function 
$\overline{S(q_x,q_\tau)}$ can
only become \emph{less} singular along the $q_x$--axis as a result of the affine
transfomation discussed above for pure systems.  Second, the slowly
decaying translational correlations which produce interesting physics
for vortex arrays in (1+1)--dimensions also appear in the ``Bragg glass''
phase which arises when the Abrikosov flux lattice is subjected to
weak point disorder in (2+1) dimensions\cite{giamarchi94}. A study of related
non-Hermitian Luttinger-liquid-like phenomena when a single twin plane
or grain boundary is inserted into bulk vortex arrays with a tilted
field and subject to point disorder is currently in progress.\cite{polkovnikov}

\section{Conclusions}
We have studied interacting vortices in a thin platelet, in the 
presence of a single columnar pin, using a combination of 
exact methods for the dilute limit based on the free fermion 
representation, field theory methods 
and numerical techniques. The response to a pin 
is controlled by the Luttinger liquid parameter, $g$, 
which depends in a complicated way on the details 
of the inter-vortex interactions and on the density. 
When the pin is parallel to the magnetic field, 
it is an irrelevant perturbation for $g>1$ but 
a relevant perturbation for $g<1$.  In both cases 
a single pin produces a local vortex lattice with 
density oscillations that decay as a power law. 
A transverse magnetic field introduces a new length scale, 
$1/h$. The density oscillations decay exponentially 
beyond this distance from the pin. We characterize 
the imaginary current, or transverse Meissner effect, 
by a ``pinning number'', $N_p$, which measures 
how many vortices are stuck in the vicinity of the pin. 
This number diverges as $(1/h)|\ln h|$ for $g<1$ 
but as $h^{2g-3}$ for $g>1$. When $g=1$, the free 
fermion mapping shows that $N_p\propto 1/h$. Even for 
zero tilt, point disorder 
drastically modified the critical behavior associated with 
$g$ passing through $1$.  However,  it may be possible to experimentally 
probe  non-Hermitian Luttinger liquid physics
using sufficiently clean high-$T_c$ thin platelets
with notches cut on the surface. 
\acknowledgements
We would like to acknowledge discussions
on the experimental situation with M. Marchevsky, P. Ong and E. Zeldov, 
conversations with L. Radzihovsky and 
helpful comments from Y. Kafri and A. 
Polkovnikov.
Work by IA was supported by NSERC of Canada and the 
Canadian Institute for Advanced Research. 
Work by WH and DRN was supported by NSF Grant  DMR-0231631 
and the Harvard Materials Research Laboratory through
NSF Grant NMR-0213805. WH was supported by a Pappalardo Fellowship. 
 WH and US ackowledge support 
from the German Science Foundation (DFG).

\appendix

\section{Critical exponents of the dilute Bose gas}
It is well-known that the low energy long distance properties of 
 a Bose gas in the dilute limit (average inter-particle 
separation, $1/n_0$, large compared to interaction range) are given by a 
gas of free fermions.  In the continuum elastic theory, defined 
in Sec. II, this 
 corresponds to the value $g=1$ 
of the dimensionless Luttinger liquid parameter  which 
determines all critical properties. In this appendix we derive a general result 
for the leading correction to this limiting value of $g$, expressing our result 
in terms of the scattering length, $a$, determined 
by the boson interaction, $V(x)$ in Eq. (\ref{Hamcon}) and the density, $n_0$. 
Apart from its applications to flux lines in thin platelets, 
we expect that this formula will be of quite general applicability 
to various one-dimensional quantum models. 

Our result depends critically on two well-known features of the Bose gas.  One 
of them is  Eq. (\ref{cg}), $cg/\pi =n_0m$. This is expected 
to be exact for a Galilean invariant gas.\cite{Haldane}  This
follows by considering the energy of the entire system lowest energy 
of momentum $P$. 
The conserved momentum density is:
\be P(x)={-i\over 2}\left[\psi^\dagger {d\over dx}\psi - \left({d\over dx}\psi^\dagger 
\right)\psi \right]\approx n_0{d\phi \over dx}.\ee
Thus the lowest energy state with a 
very low momentum  momentum $P$ is one in which $\phi (x)=(Px/n_0L)$.
From the Hamiltonian in $\phi$-representation, Eq. (\ref{H-phi}), we see that 
the energy of this state is:
\be E_0(P)={cgP^2\over 2\pi n_0^2L}.\ee
On the other hand, from Galilean invariance the exact energy of this state, which 
is simply one in which all $N$ bosons are given a boost to a momentum $P/N$ is:
\be E_0(P)={P^2\over 2mN}.\ee
Comparing these two formulas gives Eq. (\ref{cg}). We note that this argument 
doesn't really require exact Galilean invariance; it is enough that the 
disperson relation be approximately quadratic at small momentum. For instance, a quartic 
term in the dispersion relation would lead to a $(d\phi /dx)^4$ term in 
the Hamiltonian in $\phi$-representation but wouldn't interfere with 
this determination of the coefficient of the $(d\phi /dx)^2$ term. We also require
 an expression for the compressibility in terms of $g$ and $c$. 
To this end, consider the change in energy of the ground state when 
a relatively small change, $\delta N$ is made in the number of particles. From 
Eq. (\ref{n-theta}), which expresses a uniform density perturbation as
$\delta n = d\theta /dx$ and Eq. (\ref{H-u}) giving the Hamiltonian 
in $\theta$-representation,   we see that 
\begin{equation}
\delta E =  {c\pi \over 2g}{(\delta N)^2\over L}.\end{equation}
Therefore the compressibility, $\kappa$ is given by:
\begin{equation}
{1\over \kappa} \equiv {N^2\over L}\left( {\partial^2E_0
\over \partial N^2}\right)_N={c\pi \over g}n_0^2.\label{compress}\end{equation}
Upon combining Eq. (\ref{compress}) with Eq. (\ref{cg}), we see that $g$ is 
completely determined by $n_0$, $m$ and $\kappa$:
\begin{equation}
g=\sqrt{\pi^2n_0^3\kappa \over m}.\label{gk}\end{equation}
We therefore focus on calculating the compressibility $\kappa$ 
of a dilute Bose gas. 

Some general results on dilute Bose gases in (1+1) dimensions were derived in 
Ref. (\onlinecite{Lou}). There it was argued that, in 
lowest order approximation, the ground state wave-function can be 
taken to be of the form which is exact for a continuum 
$\delta$-function interaction.\cite{Lieb}  If we label 
the co-ordinates of the $N$ bosons $x_i$ and 
assume $x_1<x_2< \ldots <x_N$, this result takes the form:
\begin{equation}
\Psi (x_1,x_2,\ldots x_N)\approx \sum_PA(P)P\exp \left(i\sum_{j=1}^N
k_jx_j\right),\end{equation}
for some set of wave-vectors $k_j$ ( which must all be 
different).
 Here $P$ permutes the 
$k_j$'s and the sum is over all permutations. The $A(P)$ 
are coefficients to be determined. The many body wave-function,$\Psi$, is 
determined for other orderings of the $x_j$ from the 
required symmetry of the wave-function which follows from 
Bose statistics.  By considering what happens when 
two particles approach each other, we can see that:
\begin{equation}
A(Q)=-A(P)e^{i2\delta [(k_i-k_j)/2]}.\end{equation}
Here the two permutations $P$ and $Q$ differ only by 
interchanging particles $i$ and $j$ and $\delta (k)$
is the even channel phase shift. The quantity $\delta (k)$ is
 defined by the behavior of the 2-particle wave-function at long 
distances:
\begin{equation}
\Psi(x_1-x_2)\to \sin [k|x_1-x_2|+\delta (k)].\label{defps}\end{equation}
At small $k$, the limit which concerns us for low density, 
the phase shift behaves as:
\begin{equation}
\delta (k)\to -ak,\end{equation}
which we take as the definition of $a$, the scattering length. 
The periodic boundary conditions give a set of constraints  
which determine the allowed $k_i$'s in terms of the phase shift.  In the low 
density limit these conditions become simply:
\begin{equation}
k_j(L-Na)+a\sum_sk_s=\pi n_j,\ \  \hbox{all}\ j,
\label{k_j}\end{equation}
where the integers $n_j$ must be all even for $N$ odd 
and all odd for $N$ even. 
The solution of these equations can be further simplified 
in the low density limit, $N/L<<a$.
In lowest order we obtain simply:
\begin{equation}
k_{j0}=\pi n_j/L,\end{equation}
where the $n_j$ are  even for $N$ odd but are odd for $N$ even. 
We may calculate the  energy by simply summing the kinetic 
energy in a region of parameter space where the particles are all far apart:
\begin{equation}
E={1\over 2m}\sum_jk_j^2,
\end{equation}
so we see that we must choose the smallest possible $k_{j0}$'s to get the 
ground state corresponding to a ``Fermi surface''  for 
1D bosons. The Fermi wave-vector is determined in the usual way:
\begin{equation}
n_0=\int_{-k_F}^{k_F}{dk\over 2\pi}={k_F\over \pi}.\end{equation}
By expanding Eq. (\ref{k_j}) to next order we obtain $k_j=k_{j0}+\delta k_j$, with
\begin{equation}
\delta k_j={Na\over L}k_{j0}-{a\over L}\sum_sk_{s0}.
\end{equation}
Thus, in the limit $L\to \infty$, the ground state energy density to next order is:
\begin{equation}
E_0/L\approx {1\over 2m}\int_{-k_F}^{k_F}{dk\over 2\pi}k^2+
{a\over 2m}\int_{-k_F}^{k_F}{dk\over 2\pi}\int_{-k_F}^{k_F}
{dk'\over 2\pi}(k-k')^2.\end{equation}
Upon expressing $k_F$ in terms of $n_0$ we obtain:
\begin{equation}
E_0/L\approx {\pi^2n_0^3\over 6m}+{a\pi^2n_0^4\over 3m}.\label{E0a}\end{equation}
We can now obtain a formula for the compressibility, $\kappa$, namely 
\begin{equation}
{1\over \kappa}\equiv 
{N^2\over L}\left( {d^2E_0\over dN^2}\right)_L
=\pi^2\left({n_0^3\over m}+{4an_0^4\over m}\right).
\end{equation}
From $\kappa$ and Eq. (\ref{gk})  we determine the Luttinger liquid parameter $g$ 
at low density:
\begin{equation}
g\approx 1-2an_0 + O(a^2n_0^2).\label{glow}
\end{equation}
We note that a quartic correction to the dispersion relation would 
only correct Eq. (\ref{E0a}) and hence $1/\kappa$ at $O(n_0^5)$ and hence lead 
to a correction to $g$ of $O(n_0^2)$.  Therefore, we expect Eq. (\ref{glow}) 
to be correct even for non-Galilean invariant systems. Applications 
of this result to spin chains will be discussed elsewhere.\cite{Affleck2}

\section{determining $g$ for the tight-binding models}
Our predictions about the critical behavior of the system all involve the parameter $g$. 
To test these predictions via DMRG on the lattice tight-binding model we need to know $g$ as 
a function of the microscopic parameters of that model. At low densities 
we may obtain an analytic formula for $g$ using the general result 
Eq. (\ref{gld}) together with an exact formula for the scattering length 
of our tight binding model. For larger densities we must rely exclusively on 
numerical calculations of quantities, in the tight binding model 
with no defect and $h=0$,   
for determination of $g$. 

We now calculate the scattering length for the tight-binding 
model of Eq. (\ref{Hamlat}). 
To find the phase shift we consider the sector of the Hilbert space 
with 2 bosons, total momentum zero and even parity (as required 
by Bose statistics). 
We may write the  eigenstates 
in terms of the amplitude $\Psi_j$ for the 2 bosons to be 
separated by a displacement $j$.   
We may also write a lattice Schroedinger equation in this subspace.  
For $|j|\geq 2$ this equation reads:
\begin{equation}
-2t[\Psi_{j+1}+\Psi_{j-1}]=E\Psi_j.\end{equation}
The factor of 2 arises because we can increase the 
separation of the bosons by hopping either one of them; 
each process contributes a term $t$.  The equations
are different for $|j|=1$ or $0$ and read
\begin{eqnarray}
-2t[\Psi_2+\Psi_0]+V\Psi_1&=&E\Psi_1\nonumber \\
-2t[\Psi_{-2}+\Psi_0]+V\Psi_{-1}&=&E \Psi_{-1}\nonumber \\
-2t[\Psi_1+\Psi_{-1}]+U\Psi_0&=&E \Psi_0.
\label{Schr}\end{eqnarray}
We write the wave-function as
\begin{equation}
\Psi_j=\sin [k|j|+\delta ],\ \  (|j|\geq 1),\end{equation} 
in agreement with the definition of the phase shift in 
Eq. (\ref{defps}).  Due to the short-range interaction, this 
ansatz exactly satisfies the Schroedinger equation, 
with the correct choice of $\Psi_0$ and with $E = -4t\cos k$. 
Upon substituting in Eq. (\ref{Schr}), we obtain:
\begin{eqnarray}
-2t[\sin(2k+\delta )+\Psi_0]+V\sin (k+\delta )&=&-4t\cos k\sin (k+\delta )
\nonumber \\
-4t\sin (k+\delta )+(U+4t\cos k)\Psi_0&=&0.\end{eqnarray}
A little algebra gives:
\begin{equation}
\tan \delta ={-8t^2\sin k +(U+4t\cos k)[(V+4t\cos k)\sin k
-2t\sin 2k ]\over 
8t^2\cos k +(U+4t\cos k)[-(V+4t\cos k)\cos k
+2t\cos 2k]}.\end{equation}
Upon taking the limit $k\to 0$ we obtain $\delta \to -ak$ with:
\begin{equation}
a=-{8t^2-4tV-UV\over 2tU+UV+4tV}.\end{equation}
Note that $a$ diverges as $U$, $V\to 0$ at fixed $t$ as for the continuum 
$\delta$-function potential.  
Thus, at low density, the Luttinger liquid parameter, $g$,  behaves as:
\begin{equation}
g \to 1+2n_0{8t^2-4tV-UV\over 2tU+UV+4tV}.\label{etalow}
\end{equation}
We emphasize that this result is expected to be valid for 
arbitrarily large positive $U$ and $V$ at low enough $n_0$. 

We now turn to estimating $g$ from numerical results on finite 
systems. The most straightforward way of doing this is from 
the correlation function of $n$ or $\psi$  using Eq. 
(\ref{etadef}) or (\ref{psicorr}) respectively. Since 
our numerical results are for finite systems with periodic 
boundary conditions, we instead use the finite size 
versions of these formulas which can be obtained by a conformal
transformation:
\bea <n(x)n(0)>&\to& n_0^2 +
\hbox{constant}\times {\cos(2\pi n_0 x) \over |L\sin (\pi x/L)|^{2g}}
+\ldots \label{ncorrL}\\
<\psi^\dagger (x)\psi (0)>&\to& {\hbox{constant}\over |L\sin (\pi x/L)|^{1/2g}}
\label{psicorrL}+\ldots \eea

\begin{figure}
\includegraphics[width=8cm]{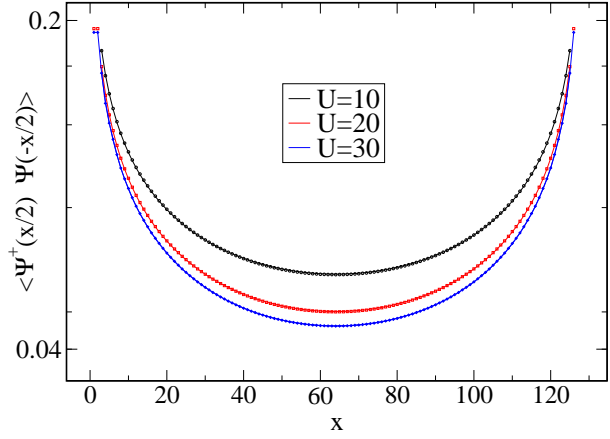}
        \caption{\label{fig:CFT_fit}
Bosonic correlation function for the lattice model (\ref{Hamlat}) 
with $L=128$, $V=0$, $n_0=0.25$ and $h=0$. 
Symbols represent DMRG results for the bosonic correlation function and  
lines are fits based on the conformal field theory prediction (\ref{psicorrL}) 
with $g=1.19/1.09/1.06$. 
}
\label{fig:psi_corr}
\end{figure}
\begin{figure}
\includegraphics[width=8cm]{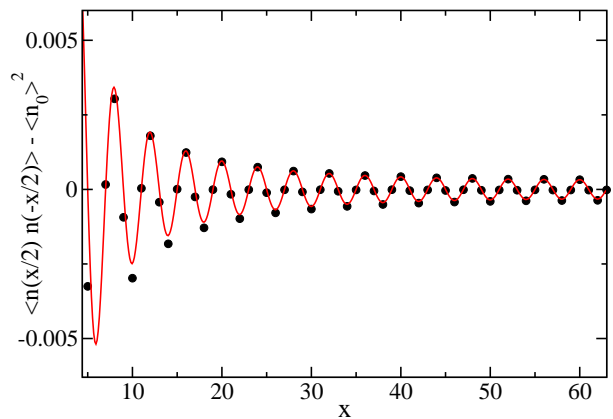}
        \caption{\label{fig:density_correlation}
DMRG result for the density correlation function with 
$U=10$, $V=4$, $L=128$, $n_0=0.25$ and $h=0$ (black circles). 
The red line is a fit based on the CFT formula (\ref{ncorrL}) with 
$g=0.72$. Note the strong Friedel oscillations due to $g < 1$.
}
\label{fig:n_corr}
\end{figure}

Alternatively, g may be extracted from the finite 
size spectrum in various ways. The compressibility is 
easily obtained from the energy to add or remove one particle:
\begin{equation}
{1\over \kappa}= {c\pi \over g}n_0^2= {N^2\over L}\left( {\partial^2E_0
\over \partial N^2}\right)_L\approx 
{N^2\over L}[E_0(N+1)+E_0(N-1)-2E_0(N)]
.\label{kappafi}\end{equation}
However, we need another result to determine separately the phonon 
velocity $c$ and the Luttinger liquid parameter $g$. A simple 
possibility is to measure the speed of sound from the 
excitation energy of the lowest excited state with wave-vector $k$, 
choosing the smallest possible non-zero wave-vector, $k=2\pi /L$. 
\begin{equation}
E(N,k)-E_0(N)\approx c|k|.\end{equation}
Another possibility is to measure the $1/L$ correction to 
the ground state energy:
\begin{equation}
E_0\approx e_0L-{\pi c\over 6L},
\label{1/L} \end{equation}
for a non-universal constant, $e_0$. This follows 
from Refs. (\onlinecite{c}) using the fact that periodic 
boundary conditions on $\psi$ correspond to periodic boundary conditions 
on $\phi$ (mod $2\pi$) or $\theta$ (mod 1).

Within our DMRG simulations we have estimated $g$ using the 
boson correlation function in eq. (\ref{psicorrL}), see  
Fig. (\ref{fig:psi_corr}). We also show the fit 
for the density correlations in Fig. (\ref{fig:n_corr}).  The 
agreement is very good except at short distances where 
the field theory predictions are expected to fail. We have checked that 
the values of $g$ so obtained are in very good agreement with the  
values obtained from the compressibility (\ref{kappafi}) 
and the $1/L$ correction to the ground state energy (\ref{1/L}).   
Note that $g$ for the Bose--Hubbard model has been calculated previously  
by K\"uhner et al. \cite{Kuehner} where the authors used 
periodic boundary conditions in the DMRG algorithm. 
Our results for $g$ are shown in fig.~(\ref{fig:eta_vs_density}). 
For small densities the agreement between the asymptotic expression 
(\ref{etalow}) and the numerical data is obviously very good. 
We expect that with increasing density, $g$ behaves in a non-monotonic 
fashion and finally approaches $g=1$ again as the filling becomes 
commensurate in the limit $n_0\to 1$. 
For finite next--neighbor repulsion $V>0$ and $n=0.5$ a charge--density 
wave instability occurs\cite{Kuehner} which we have not studied here. 
For the main motivation of our work -- interacting vortex physics -- 
the Bose--Hubbard model is only applicable in the limit of small filling, 
since the lattice constant has no direct physical meaning. 
Most of the results in our paper have been obtained 
for $n_0=0.25$.

\begin{figure}
\includegraphics[width=8cm]{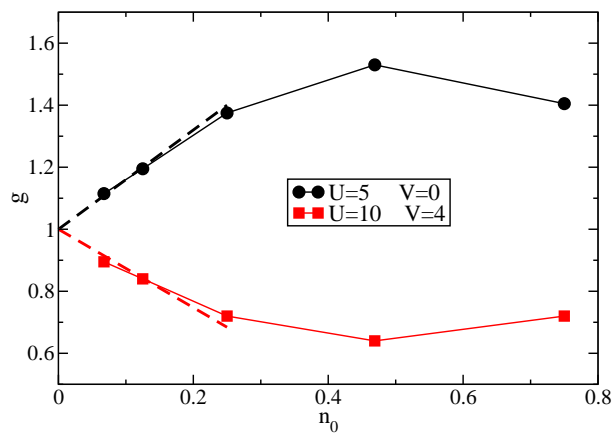}
        \caption{\label{fig:eta_vs_density}
Luttinger--liquid parameter $g$ calculate for the lattice model (\ref{Hamlat}) 
with $h=\epsilon_0=0$ 
as a function of the boson density $n_0$. 
We have plotted results for two different values of the interaction parameters 
corresponding to $g>1$ and $g<1$, respectively. 
The dashed lines show the analytic result (\ref{etalow}) at low densities. }
\end{figure}

\section{boson correlation function with a pin}
In this appendix we discuss numerical and analytic results on
the correlation function of the boson creation operator across a pin, 
in the absence of a tilt field, $h=0$. This correlation function 
clearly distinguishes the cases of a relevant ($g<1$) and irrelevant ($g>1$) pin. 

We note that the finite $L$ equal time boson correlation function 
at the broken chain ($\Gamma \to 0$) fixed point, for $x,y>0$, is given by 
 a conformal transformation of Eq. (\ref{psicorimp}): 
\be <\psi^\dagger (x)\psi (y)>\propto 
\left\{\sin (\pi x/L)\sin (\pi y/L)
\over L^2\sin^2 [\pi (x-y)/2L]\sin^2 [\pi (x+y)/2L]\right\}^{1/4g}.
\label{psicorN}\ee
This correlation function is now small but non-zero for two 
points near but on opposite sides of the pin, $x<<L$ and $L-x<<L$. These
points are correlated by going around the chain, rather than 
across the pin. 
Even for finite $\ep$ we expect this result to hold in the case 
of a relevant pin, $g<1$, at sufficiently large $L$, $x$, $y$, $N-x$ 
and $L-y$.  We note that neither $x$ nor $y$ can be too close to 
the pin.

We now consider specifically the boson correlation function {\it across} the pin, 
$<\psi^\dagger (x)\psi (-x)>$, ($x>0$) in the limit of 
infinite system size. If the pin is irrelevant, $g>1$, then 
at large $x$ we should recover the result for the system with no 
pin:
 \be <\psi^\dagger (x)\psi (-x)>\to {\hbox{constant} \over x^{1/2g}}.\ee
  On the 
other hand, if the pin is relevant, $g<1$, this correlation function 
vanishes more rapidly with $x$. The weak tunnelling, $\Gamma$, model 
is again useful for this calculation. We may calculate 
this correlation function in lowest (first) order perturbation theory 
in $\Gamma$ and then ``renormalization group improve'' the calculation 
by replacing $\Gamma$ by its renormalized value at scale $x$. Perturbation theory  
gives:
\be <\psi^\dagger (x)\psi (-x)>\to -\Gamma \int d\tau 
<\psi^\dagger (x,0)\psi (0^+,\tau )>_0<\psi^\dagger (0^-,\tau )\psi (-x,0)>_0
+c.c.\label{psiacint}\ee
The correlation functions inside the integral, evaluated using the 
Hamiltonian  of Eq. (\ref{Heff}) with $\Gamma =0$, may be obtained 
from Eq. (\ref{psicorimp}) by setting $y$ to a value of order 
a short distance cut off:
\be <\psi^\dagger (x,0)\psi (0^+,\tau )>_0\propto 
\left\{x\over [x^2+c^2\tau^2]^2\right\}^{1/4g}
\label{boscor0}\end{equation}
Upon inserting this expression into the integral of Eq. (\ref{psiacint}), we obtain
\be <\psi^\dagger (x)\psi (-x)>\propto  {\Gamma \over x^{3/2g-1}}.\ee
From Eq. (\ref{GammaRG}) we see that
\be \Gamma_{\hbox{eff}}(x)\propto {1\over x^{1/g-1}},\ee so finally
\be <\psi^\dagger (x)\psi (-x)>\propto {1\over x^{5/2g-2}}.\label{psiac}\ee
Since we are assuming now that $g<1$, we see that this 
correlation function drops off 
more rapidly than $1/x^{1/2g}$, the result for no pin.

Finally, if we take into account the finite $L$ periodic boundary 
conditions in the case of a relevant pin, 
$<\psi^\dagger (x)\psi (-x)>$ would be given by a 
sum of the result for an infinite pin of Eq. (\ref{psicorN}) and 
the result of Eq. (\ref{psiac}), suitably generalized to take 
into account the finite $L$. For $x<<L$, this is the sum of two 
small terms.  One is the weak correlations across the pin and 
the other is the correlations from going around the circle the long way 
without crossing the pin.

Numerical results for the boson correlation function are shown 
in fig. \ref{fig:correlation_w_impurity}. 
Evidently, correlations across the pin are strongly suppressed if the 
defect is relevant ($g<1$). From the numerics it is unclear 
whether the exponent of the power law decay changes as 
given by (\ref{psiac}).
On the other hand, a pin of the same strength 
has almost no effect if $g>1$.

\begin{figure}[h]
\includegraphics[width=8cm]{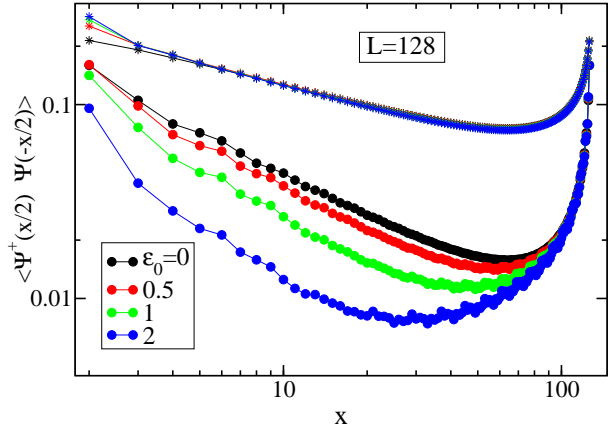}
        \caption{\label{fig:correlation_w_impurity}
Impurity effect on the bosonic correlation function. 
For a relevant pin with $U=10$, $V=4$ and $g=0.72$ (circles) we find 
a strong suppression of correlations close to the impurity 
(which is at $x=0$).
On the other hand, for an irrelevant pin with 
$U=5, V=0$, $g =1.41$ (stars) the dependence 
on the scattering potential $\epsilon_0$ is very weak. (These
are the upper curves which lie almost on top of each other.)
}
\end{figure}

\section{periodic/anti-periodic ground state energy difference}

In this appendix we discuss the difference in 
ground state energies with anti-periodic and periodic boundary conditions 
on $\psi$.  In the tight-binding model this corresponds to changing 
the sign of the hopping term on any single link. From Eq. (\ref{bospsi}) 
we see that anti-periodic boundary conditions on $\psi$,
\begin{equation}
\psi (x+L)=-\psi(x),\end{equation}
imply that the phase field $\phi$ must wind by a half-integer:
\begin{equation}
\phi (x+L)=\phi (x)+(2p+1)\pi,\label{bcAP}\end{equation}
for integer $p$. The ground states correspond to $\phi (x)=\pm \pi x/L$ 
with energy:
\begin{equation}
E^{AP}_0-E^P_0={\pi cg\over 2L}.\label{EPAP}\end{equation}
Upon noting that Eq. (\ref{EPAP}) determines $cg$ while the compressibility gives 
$g/c$, we see that both parameters are then determined. 

The effects of a pin on the periodic/anti-periodic ground state energy 
difference can  be calculated using ``renormalization group 
improved'' perturbation theory. Antiperiodic boundary conditions 
in the tight-binding model 
with a pin are defined by the Hamiltonian of Eq. (\ref{Hamlat}) but 
with the periodic condition of eq. (\ref{bc}) replaced by:
\be b_L\equiv -b_0.\label{APbc}\ee
When the pin is irrelevant, $g>1$ and we expect to obtain the
result of Eq. (\ref{EPAP}) asymptotically for large $L$. On the other hand, 
if the pin is relevant, $g<1$,
the two sides of the system become asymptotically decoupled and 
 this energy difference scales 
to zero more rapidly with $L$. Note that $E^{AP}_0-E^P_0$
 is strictly zero at the open chain fixed point. Within 
the effective Hamiltonian of Eq. (\ref{Heff}), we see that when $\Gamma =0$, 
it makes no difference whether we impose a boundary condition 
$\psi (0^+)=\pm \psi (0^-)$. For small $\Gamma$, this energy 
difference is first order in $\Gamma$.  From Eq. (\ref{psicorN}), 
with $x$ and $y$  of order a short distance cut off, we obtain:
\be E_0^{AP}-E_0^P\propto {\Gamma \over L^{1/g}}.\ee
If we now replace $\Gamma$ by its renormalized value at scale $L$, we obtain:
\be E_0^{AP}-E_0^P\propto {1\over L^{2/g-1}}.\ee
Since $g<1$ here, we see that the decay is always faster than with no pin. 

In fig. (\ref{fig:BC_sensitivity}) we show numerical results for the 
boundary sensitivity $\Delta E_0 \equiv E_0^{AP} - E_0^{P}$ 
normalized to the value without any defect. Although our system 
sizes were not big enough to extract power law behaviors, 
the data suggest that for a relevant pin ($g<1$) 
the normalized sensitivity decreases to zero in the 
thermodynamic limit, indicating that the system is indeed asymptotically 
``cut in two''. (For a very large system, we expect
$\Delta E_0(\hbox{impurity}/\Delta E_0(\hbox{no impurity})\propto L^{2-2/g}$.)
For $g>1$, on the other hand, our numerical results are consistent 
with a ratio which 
 approaches unity for large $L$, indicating the irrelevance of the pin 
at large length scales. 

\begin{figure}[h]
\includegraphics[width=8cm]{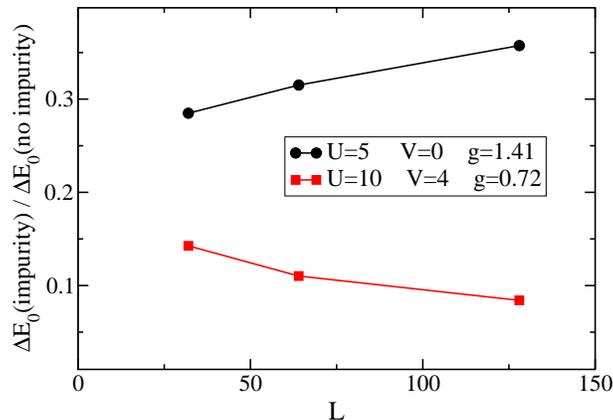}
        \caption{\label{fig:BC_sensitivity}
Sensitivity of the ground state energy to 
a change from periodic to antiperiodic boundary conditions 
for $n_0=0.25$ and $\epsilon_0=4$. 
}
\end{figure}

The Hamiltonian version of the $\phi$-representation is 
convenient for studying the periodic/anti-periodic ground state 
energy difference in the presence of a tilt field, $h$ (but no pin). 
[See Eqs. (\ref{Hamcon}) and (\ref{Hphih}).]
As discussed 
above Eq. (\ref{bcAP}),  for anti-periodic boundary conditions, 
$\phi$ must wind by $\pm$ half a period in the ground state:
$\phi (x)=\pm \pi x/L$.  Note that the non-Hermitian term does 
not change the ground state but simply adds an imaginary term to its 
energy. We now have:
\be E_0^{AP}-E_0^P=\pm {ihn_0\pi\over m} +\pi cg/2L.\ee
The imaginary term is $O(1)$, unlike the real term which is $O(1/L)$.

\section{Strength of point disorder}
In this appendix we estimate the conditions such that the effect of residual point disorder 
on the thermally excited vortex lines, which are the 
main subject of this paper, can be ignored. This neglect can be justified only 
over a limited range of length scales. As discussed in Sec. 5, point disorder 
eventually changes the physics for any value of $g$.  
Our analysis is mostly based on Refs.~(\onlinecite{nelson92}) 
and (\onlinecite{Lehrer00}), to which we refer readers for 
additional details.

The discussion proceeds in two stages. First, we demonstrate that 
in (1+1)--dimensional vortex arrays, typical pinning energies per unit length 
associated with an isolated columnar pin or ``notch'' greatly 
exceed the energy available from ignoring the pin and simply following 
an optimal path through the point disorder. Thus the columnar 
pinning energy greatly exceeds the collective effect of point pins. 
Second, given that at least one vortex line is strongly pinned on a 
columnar defect, we provide a rough estimate of a sufficient condition 
for thermal fluctuations of vortices to dominate over 
point disorder--induced wandering until one reaches the length scale 
of the renormalization group equations of Sec. 5. 
Our main conclusion is that (1+1)--dimensional ``platelet'' samples 
of high--$T_c$ materials 
(disordered only by oxygen vacancies) within 5-10\% of the critical temperature 
could provide a good opportunity to study the physics of thermally excited 
vortex lines discussed in this paper over a range of length scales. 

We first compute the average pinning energy of a single vortex, confined by its 
neighbors in a (1+1)--dimensional array, due to random point impurities. 
We define a characteristic ``imaginary time'' length scale $l^*$ at which 
transverse fluctuations of a vortex line due to point disorder reach the 
average vortex spacing $a_0$: 
\be
\delta x(l^*) = x_c \left(\frac{l^*}{l_c}\right)^\zeta = a_0 
\label{disorder_fluct} 
\ee 
where $\zeta = 2/3$ in (1+1)--dimensions\cite{Lehrer00} and  
\be
x_c = T^3/\tilde{\epsilon}_1 \tilde \Delta_0,
\label{thermal_length}
\ee
\be
l_c = \frac{\tilde{\epsilon}_1}{T}\, x^2_c.
\ee 
Here $x_c$ is the length scale out to which vortex wandering is described 
by a thermal random walk, and $l_c < l^*$ is the 
corresponding length in the imaginary time direction. 
Note that $\tilde{\epsilon}_1$ is the line tension introduced  
in (\ref{eq:first}) and $\tilde \Delta_0$ is a correlator 
measuring the strength of point disorder.  The 
parameter $\tilde \Delta_0$ is related to the correlator 
$\Delta_0$ introduced in the continuum model of Eq. (\ref{disorder_correlator}) by 
$\tilde \Delta_0=\Delta_0/n_0^2$.   
The pinning energy for an imaginary--time 
segment of length $l$ on this path is given by\cite{Lehrer00} 
\be
U_p(l^*) = T \left(\frac{l^*}{l_c}\right)^{2\zeta - 1}. 
\ee
Upon solving for $l^*$ and using $\zeta=2/3$ the 
pinning energy per length due to point disorder simplifies to:
\be
\frac{U_p(l^*)}{l^*} = \frac{\Delta_0}{a_0 T}.
\ee
For quantitative estimates, we assume isotropy of the effective mass in the 
$ab$--plane and average over the sample thickness along the $c$--axis to obtain 
the disorder correlator as\cite{nelson92} 
\be
\Delta_0 = \frac{1}{w} \left(\frac{\phi_0}{4\pi \lambda}\right)^4 
\, \xi^3 \, \left(\frac{J_{cp}}{J_{pb}}\right)^{3/2}, 
\ee
where $J_{pb}$ is the pair breaking critical current, $J_{cp}$ is the critical 
current due to point defects and $\phi_0 = 2\times 10^{-7} G cm^2$ is the flux quantum. 
With sample thickness $w=150$ nm, 
penetration depth $\lambda = 150$ nm, coherence length 
$\xi = 2$ nm, temperature $T=50$ K, and the estimate 
(valid for high--$T_c$ superconductors\cite{blatter94})
 $J_{cp}/J_{pb} \approx 0.01$,   
we obtain a pinning energy per length 
\be
\frac{U_p(l^*)}{l^*} \approx 7\times 10^{-12} \hbox{erg/cm}. 
\ee
For comparison, the pinning energy per unit length due to a columnar defect 
is given by\cite{nelson92}
\be
U_0/L = \left(\frac{\phi_0}{4\pi \lambda}\right)^2 \approx 
1.2\times 10^{-6} \hbox{erg/cm}
\ee
and is thus clearly dominant compared to point disorder by several orders 
of magnitude. 

A related question concerns whether disorder affects the wandering 
of a \emph{single} vortex line in (1+1)--dimensions before it 
interacts with its neighbors. Physically, we require that thermally 
excited vortices collide and restart their random walk 
several times before 
pinning due to point disorder becomes strong. 
Up to constants of order unity, the requirement is that the 
thermal length scale $x_c$ in Eq.~(\ref{thermal_length}) above exceed  
the vortex spacing $a_0$, 
\be
x_c = \frac{T^3}{\tilde{\epsilon}_1 \Delta_0} > a_0
\ee
Using the same parameters as above, we find 
\be
x_c = 3.5\times 10^{-7}cm  
\ee
at $T=50 K$. Upon assuming that $J_{cp}/J_{pb}$ is roughly temperature 
independent and noting that close to the critical temperature  
$\tilde{\epsilon}_1 \Delta  \propto  \lambda^{-3}(T) \approx \lambda^{-3}(50 K) |t|^{3/2} $ 
we have 
\be
\frac{T^3}{\tilde{\epsilon}_1 \Delta} = 
\stackrel{\approx 3.5\times 10^{-6}cm}
{\overbrace{\frac{T^3}{\tilde{\epsilon}_1 \Delta}\Big|_{T=50 K}}} \, 
\left(\frac{T}{50 K}\right)^3 \, |t|^{-3/2}. 
\ee
Here 
\be t\equiv (T-T_c)/T_c.\ee
For a typical high-$T_c$ compound like YBCO with $T=80 K$ and a reduced 
temperature $|t|= |T-T_c|/T_c = 0.1$ 
we thus obtain 
\be
x_c \approx 2.7\times 10^{-5}cm > a_0
\ee
which is larger than a typical vortex spacing 
for $H \gtrsim \phi_0/(w x_c) \approx 400 G$.
 With these estimates in hand, we can use renormalization group recursion 
relations [see Sec. 5 and Ref. (\onlinecite{polkovnikov})] to determine 
the length scale at which point disorder dominates. For clean 
high $T_c$ samples close to the critical point, this  length scale can 
be very large.

\section{Connection with string theory}
The bosonized form of the weak tunnelling model, Eq. (\ref{Heff}), 
generalized to arbitrary $g$,  can be simplified 
by introducing linear combinations of the (independent) boson fields to the left and 
right of the pin:
\be \phi_{\pm}(x) \equiv [\phi (x)\pm \phi (L-x)]/\sqrt{2}.\ee
Despite the non-local nature of this transformation, it yields a local 
Lagrangian since the interaction (tunneling) term occurs only at $x=0$. 
This Lagrangian takes the form:
\be L=\sum_{\pm}\int_0^{L/2} dx\left\{{g\over 2\pi}\left[
{1\over c}\left({\partial \phi_{\pm}\over \partial \tau}\right)^2
+c\left({\partial \phi_{\pm}\over \partial x}\right)^2\right] 
+{ihcg\sqrt{2}\over \pi}{\partial \phi_-\over \partial x}\right\}
+\tilde \Gamma \cos [\sqrt{2}\phi_-(0)]
\ee
where $\tilde \Gamma \propto \Gamma$, the weak tunnelling amplitude 
introduced in Eq. (\ref{Heff}). 
We see that $\phi_+$ decouples from both the pin and the 
tilt field. $\phi_-$ obeys mixed boundary conditions,
$\phi_- (L/2)=0$, $d\phi_-/dx (0)=0$.

Alternatively, we may make the similarity transformation 
of Eq. (\ref{NUT}) {\it before} rewriting the Lagrangian in terms of 
the phase field, $\phi$. 
This is equivalent to:
\be \phi (x) \to \phi (x)+ihx,\ee 
implying:
\be \phi_-(x)\to \phi_-(x)+ih(2x-L)/\sqrt{2}.\ee
This eliminates the term proportional 
to $ih\partial \phi_- /\partial x$ but makes the tunnelling 
term non-Hermitian.  Dropping the field $\phi_+$, we obtain 
the equivalent Lagrangian:
\be  L={g\over 2\pi}\int_0^{L/2} dx\left[
{1\over c}\left( {\partial \phi_{-}\over \partial \tau}\right)^2
+c\left({\partial \phi_{-}\over \partial x}\right)^2\right] 
+{\tilde \Gamma \over 2}\left[ e^{hL}e^{i\sqrt{2}\phi_-(0)}
+e^{-hL}e^{-i\sqrt{2}\phi_-(0)}\right].
\ee
In the small $\tilde \Gamma$ limit, it is permissible to 
drop the exponentially small term, $\propto \tilde \Gamma e^{-hL}$, 
leaving the Lagrangian:
\be  L={g\over 2\pi}\int_0^{L/2} dx\left[
{1\over c}\left( {\partial \phi_{+}\over \partial \tau}\right)^2
+c\left({\partial \phi_{+}\over \partial x}\right)^2\right] 
+{\tilde \Gamma \over 2} e^{hL}e^{i\sqrt{2}\phi_-(0)}.
\ee
Finally, if we make the formal analytic continuation, 
$\phi_-(x,\tau )\to -i\phi_-(x,\tau )$
we obtain:
\be  L=-{g\over 2\pi}\int_0^{L/2} dx\left[
{1\over c}\left( {\partial \phi_{-}\over \partial \tau}\right)^2
+c\left({\partial \phi_{-}\over \partial x}\right)^2\right] 
+{\tilde \Gamma \over 2} e^{hL}e^{\sqrt{2}\phi_-(0)}.
\ee
This is a strong coupling limit of a 
model, known as the time-like boundary Liouville 
theory, which has received considerable attention in the 
string theory literature lately (in the marginal 
case, $g=1$), because of its connection 
with ``s-branes'', i.e. space-like ``d-branes''.
See, for example, Ref. (\onlinecite{Gutperle}).

\end{document}